# Interchromophoric Interactions Determine the Maximum Brightness Density in DNA Origami Structures

*Tim Schröder[1], Max B. Scheible[2], Florian Steiner[1], Jan Vogelsang[1,*], Philip Tinnefeld[1,*]*

1. Department Chemie and Center for NanoScience, Ludwig-Maximilians-Universitaet Muenchen, Butenandtstr. 5-13 Haus E, 81377 Muenchen, Germany.

2. GATTAquant GmbH, Am Schlosshof 8, 91355 Hiltpoltstein

ABSTRACT

An ideal point light source is as small and as bright as possible. For fluorescent point light sources, homogeneity of the light sources is important as well as that the fluorescent units inside the light source maintain their photophysical properties which is compromised by dye aggregation. Here we propose DNA origami as a rigid scaffold to arrange dye molecules in a dense pixel array with high control of stoichiometry and dye-dye interactions. In order to find the highest labeling density in a DNA origami structure without influencing dye photophysics we alter the distance of two ATTO647N dyes in single base pair steps and probe the dye-dye interactions on the single-molecule level. For small distances strong quenching in terms of intensity and fluorescence lifetime is observed. With increasing distance, we observe reduced quenching and molecular dynamics. However, energy transfer processes in the weak coupling regime still have a significant impact and can lead to quenching by singlet-dark-state-annihilation. Our study fills a gap of studying the interactions of dyes relevant for superresolution microscopy with dense labeling and for single-molecule biophysics. Incorporating these findings in a 3D DNA origami object will pave the way to bright and homogeneous DNA origami nanobeads.

Fluorescence enables ultra-sensitive detection down to single molecules. Nevertheless, the signal emitted from single molecules is often too weak to be detected with simple devices and in complex environments. As single organic dye molecules also suffer from photobleaching the overall obtainable signal is too small for many applications and larger, often multi-chromophoric alternatives are required.[1] Brighter alternatives such as dye loaded nanoparticles and quantum dots are larger and more heterogeneous limiting their usefulness as point light sources.[2-4] In optical characterizations such as the determination of a point-spread-function of a STED microscope,[5] the small size of the light source is central. Similarly, optical setup characterizations of, for example, sensitivity require homogeneous, well characterized light sources. Moreover, for bio-molecular imaging and molecular diagnostics the size of the light source negatively influences biocompatibility as well as diffusion and binding kinetics. Quantum dots have additional disadvantages of toxicity and size in bio-applicable formulations.

Bright fluorescent particles are commonly produced by embedding fluorescent dyes in a polymer particle of nanoscale dimensions. The density of dyes that can be embedded is limited by the dye's tendency to aggregate yielding low-fluorescent imperfect H-type aggregates.[6-10] Often it is also observed that dye interactions lead to broad brightness and fluorescence lifetime distributions especially for nanobeads below 60 nm diameter.[11] Only recently, fluorescent nanoparticles with improved photophysical properties such as conjugated polymer dots[12, 13] and fluorescent organic nanoparticles[4, 14] were developed.[15] Still, these nanoparticles exhibit substantial heterogeneity in size and shape as well as limited control over surface chemistry. Moreover, these nanoparticles exhibit a limited linear dynamic range of emission versus excitation power and their brightness could yet also not be referenced to that of single dye molecules.

Here, we investigate DNA origami as scaffold for ultra-bright point light sources. In DNA origami, a roughly 7000-8000 nucleotides long, single-stranded DNA strand is folded into a programmed three-dimensional shape by ~200 staple strands that have a length of the order of 32 nucleotides.[16, 17] In DNA origami nanobeads, the DNA nanostructure provides a three-dimensional scaffold to which dye molecules can be attached in a pre-defined pattern. This pre-defined pattern could be a regular, three- or two-dimensional arrangement in which the dye molecule positions are representing voxels.

If staple strands are labeled at the end with a single dye molecule, usually a dye arrangement with inter-dye distance of ~6 nm can be obtained for a two-dimensional DNA origami structure as the one indicated in Figure 1. With these 6 nm-pixels, dye molecules can interact via Förster Resonance Energy Transfer,[18] but the brightness and fluorescence lifetime of identical fluorescent dyes is not affected.[11, 19] We note that the distance of 6 nm can be decreased by redesigning the 2D or 3D DNA origami structure or by considering additional internal labeling of the staple strands. Therefore, we set out to identify the highest brightness density that can be achieved with DNA origami and the optimal distance of placing fluorescent dyes, here ATTO647N, without quenching. Using single-molecule spectroscopy, we reveal five distinct interaction processes and dynamics that have to be considered to satisfactorily describe the apparently simple two-dye system in DNA origami.

To study the distance dependence of the fluorescent properties of two dye molecules we placed them in a two-dimensional rectangular DNA origami model structure, as schematically depicted in Figure 1 (see supporting information for AFM pictures in Figure S1 of the DNA origami, sample preparation and methods). We altered the distance of these two red emitting dyes on one helix in single-nucleotide steps. Green ATTO542 dyes were also incorporated into the DNA origami structure for identifying DNA origami structure locations. Single-molecule measurements were performed on a BSA-biotin coated glass surface (see supporting information for details). The signals were analyzed in terms of fluorescence intensity and fluorescence lifetime for each single DNA origami structure.

For placing the dye molecules in the DNA origami structure, we used commercially available dye chemistry. The first dye was attached at the 5'-end via a C6 amino linker to an oligonucleotide, which was used in every DNA origami structure in this study. To keep the environment similar, the second dye was attached to the next oligo at the 3'-end by a C7 amino linker. For each distance a different labeled oligo was used in the folding process. At the smallest possible distance, both dye-labeled ends face each other. This distance is referred to as 1 base pair (bp). For greater distances, e.g. 3 bps (the 3' labeled oligo is two nucleotides shorter than the 1 bp oligo), we filled the two unpaired bases at the scaffold with a complementary sequence attached to an oligo from the neighboring helix, as is shown in the magnified view in Figure 1. Forming dsDNA should rigidify the structure. 60 nm away from the ATTO647N dye-couple we placed up to 10 ATTO542 dyes by external labeling to identify the location of DNA origami structures on the coverslip surface.[19]

The confocal fluorescence microscopy surface scans in Figure 2a with alternating laser excitation at 532 nm and 640 nm show strong intensity quenching for small (1–3 bp) distances between the red dyes (see supporting information for details on the setup). Dark pixels inside the diffraction-limited spots for the intermediate distances (4–6 bps) reveal pronounced blinking. For the 3 bp- to 7 bp-distance samples we introduced a so-called spacer strand, which binds to the single stranded scaffold domain in between the two dyes to separate them by building a double helical filling (see Figure 1). The spacer is a ssDNA-strand, which originates from the neighboring helix and binds into the gap between the two dyes. For the 3 bp-distance sample, the spacer strand has therefore only two nucleotides to form a duplex, which is thermodynamically unstable at room temperature. Although the spacer strand has a high local concentration by being fixed to the adjacent helix – which should push the equilibrium towards the separation of the two dyes – the introduced spacer strand has no influence at the 3 bp-distance sample due to the weak hybridization energy of only two nucleotides. At greater distances (4 bps and 5 bps), a stronger fluorescence signal was obtained with pronounced blinking. At 6 bps the blinking is largely vanished. Only a few spots demonstrate blinking. The blinking vanishes completely at 7 bps distances indicating permanent separation of the dyes. Figure S2 in the supporting information further demonstrates that the blinking kinetics are slowed down by increasing the length of the spacer strand. The comparison with the dual color image, i.e. detecting ATTO542 and ATTO647N simultaneously, shows a perfect match of green and red spots without blinking. The assignment of the blinking to the hybridization and dehybridization of the spacer strand from the neighboring helix is confirmed by measurements without the spacer extension. In this case the DNA linker between the dyes is so floppy that quenching can even occur up to 10 nm (see Figure S3).

The adjustable blinking between an unquenched and a quenched state of the bi-chromophoric system allowed us to directly compare these two states with each other at the single-molecule level to reveal the nature of the quenched state. We recorded fluorescence transients as shown in Figure 2b for the 5 bp-distance sample under oxygen removal and reducing and oxidizing (ROXS) conditions, thus blinking due to photophysics of the dyes themselves is successfully suppressed.[20] Hence, the fluorescence blinking can be solely attributed to dye-dye-interactions. Two fluorescence intensity levels are observed, a high-intensity state with a mean intensity of 61 ± 12 kHz (marked in red) and a low-intensity state with 7 ± 3 kHz (marked in black). The background level (marked in blue) is below 0.5 kHz. We note that the high-intensity level corresponds to approx. two-times the intensity of a single dye. Additionally, the fluorescence lifetimes $\tau_{Fl,high}$ and $\tau_{Fl,low}$ of the high- and low-intensity states are recorded by time-correlated single photon counting (TCSPC), respectively, yielding $\tau_{Fl,high} = 4.5 \pm 0.1\ ns$ and $\tau_{Fl,low} = 3.6 \pm 0.1\ ns$. The fluorescence lifetime of the high-intensity state corresponds to the lifetime of a single ATTO647N reference dye. As the high-intensity state resembles two completely unquenched ATTO647N dyes with a reported fluorescence quantum yield $QY_{Fl} = 65\ \%$ and assuming that the low-intensity state is solely attributed to a drop in $QY_{Fl}$ (as both two-times the monomer and dimer spectra show similar extinction at the excitation wavelength of 637 nm, see Figure 3c) we could estimate the radiative, $k_r$, and non-radiative, $k_{nr}$, rate constants of both states.

For the high-intensity state we calculated $k_{r,high} = 1.4 \cdot 10^8 s^{-1}$ and $k_{nr,high} = 0.77 \cdot 10^8 s^{-1}$, whereas the low-intensity state yielded $k_{r,low} = 0.2 \cdot 10^8 s^{-1}$ and $k_{nr,low} = 2.6 \cdot 10^8 s^{-1}$. In conclusion, the radiative rate drops down by almost an order of magnitude whereas the non-radiative rate increases by a factor of ~3.5 in the quenched state.

For a statistical analysis regarding the change of rate constants, we extracted the fluorescence intensity and lifetime of each diffraction limited spot. We note that the integration time per measured spot is approximately 100 ms and, therefore, the measured fluorescence lifetime is a temporal average. More than 400 spots were analyzed per sample and plotted in a scatter plot. Figure 2c shows weak fluorescence combined with a shortened fluorescence lifetime for the 1 bp- to 3 bp-distance samples. The inset of Figure 2c demonstrates the decrease of the mean fluorescence brightness with decreasing distance between both dyes. We simplified that 1 bp corresponds to ~ 0.34 nm, neglecting the impact of the dsDNA backbone on the distance. The fluorescence intensity is homogenously ten times lower than expected for two unquenched dyes while the fluorescence lifetime is shorter by a factor of two compared to the unquenched 6 bps- to 8 bps-distance samples. The medium distances with 4 and 5 bps display a broad distribution, essentially connecting the unquenched and quenched populations. By extracting the mean brightness and fluorescence lifetime for the quenched populations, i.e. the 1-3 bps-distance samples, and the unquenched populations, i.e. 6-8 bps-distance samples, we can estimate the following representative mean rate constants for both populations: $\langle k_{r,high} \rangle = 1.5 \cdot 10^8 s^{-1}$, $\langle k_{nr,high} \rangle = 0.8 \cdot 10^8 s^{-1}$, $\langle k_{r,low} \rangle = 0.3 \cdot 10^8 s^{-1}$ and $\langle k_{nr,low} \rangle = 3.9 \cdot 10^8 s^{-1}$.

Both changes of $k_r$ and $k_{nr}$ unravel that two quenching mechanisms must be responsible for the strong decrease of the fluorescence, when the two dyes come too close to each other. The almost 3-fold decrease of $k_r$ can be associated with an H-type dimer formation, i.e. an imperfect H-aggregate, which is still fluorescent. A cofacial stacking of both dyes leads to a splitting of the excited state. Further, the transition dipole moment (TDM) of the higher lying excited state is enhanced whereas the TDM of the lower excited state is significantly reduced. For these reasons, the excitation of such a dimer is shifted to higher energies, known as a hypsochromic shift, which can be seen in the absorption spectrum of a comparable dsDNA model structure, which is shown in Figure 3. Further, the fluorescence from the lower lying state is slowed down. An absorption spectrum requires roughly 30 batches of DNA origami structures to reach the needed amount of substance. For this reason, we choose dsDNA as a model for two extreme distances of 1 bp and 20 bps. The model structures are shown in Figure 3a. Dye labeled oligo sequences were maintained. Other distances need a spacer between the two dyes and short nucleotides are thermally not stable. In the DNA origami structure a neighboring helix is present where the spacer oligo is attached, which lead to a high local concentration of the separating sequence when it is in an unbound state. The dsDNA 1 bp-distance model structure matches the intensity and fluorescence lifetime distribution of the 1 bp-distance DNA origami structure, when immobilized on the surface as shown in Figure 3b, which justifies the comparison. The absorption spectrum of the dsDNA 1 bp-distance model structure shows the hypsochromic shift at 605 nm (Figure 3c). Similar results were obtained for cofacially stacked pi-conjugated oligomers by Stangl *et al.*[7] However, the major difference here is given by the environment. Whereas the dimers in the work by Stangl *et al.* were embedded in poly(methylmethacrylate), i.e. solid state, we have a dynamic solution environment. The dimers can undergo rapid conformational changes due to the C6 and C7 linkers of the dyes allowing for rotational diffusion. Such freedom leads to rapid collisions between both dyes, hence dynamic quenching and an increase of $k_{nr}$. The change of $k_{nr}$, $\Delta k_{nr} = 1.83 \cdot 10^{8} s^{-1}$, corresponds to the collision frequency and relates to one collision per 5.5 ns, a similar time regime in which rotational diffusion of dyes attached to DNA takes place.[21]

Intensity and fluorescence lifetime distributions become very broad for the 4 bps- and 5 bps-distance samples. Low intensity always correlates with shortened fluorescence lifetime but it does not scale linear but in a mirrored/reverse “L” shape (Figure 2c). This can be explained by the spacer separating the two dyes. The dyes are well separated while the spacer is hybridized to the scaffold (unquenched state) and therefore we observe the expected intensity and fluorescence lifetime of two dyes. But the sequence of the separation spacer is only 3 and 4 nt long and is not thermally stable. When the separating part of the spacer dissociates the dyes are no longer separated and they can quench each other by the static and dynamic quenching mechanisms discussed above.

In the next step, we focus on the 7 bp-distance sample, for which the spacer strand is stably incorporated. 7 bps between both dyes corresponds to a distance of ~2.4 nm which is typical for weak coupling effects, i.e. energy transfer in the Förster regime (FRET). Whereas FRET between identical dyes (Homo-FRET) might occur it has no influence on the emitted fluorescence intensity at the here used excitation intensities as the emissive rate is independent of which of the identical dyes is in the excited state. Still, energy transfer processes might occur that limit the usefulness of dense labeling in the Förster regime through saturation of the emitted fluorescence.[22] Such energy transfer processes can lead to quenching of excited singlet states of one dye by light-absorbing states of other dyes in close proximity. These light-absorbing states can be associated with excited singlet states, $S_1$,[23] excited triplet states, $T_1$,[24, 25] or radical states.[26] To test whether such quenching mechanisms play an important part for the here investigated system we analyzed the photon stream of the 7 bp-distance sample by photon correlation techniques and compare this sample with a single chromophore sample and the 20 bp-distance sample.

First, we investigate, if quenching of excited singlet states by another chromophore in its excited singlet state, i.e. singlet-singlet annihilation, does play a significant role, which will lead to single photon emission.[23, 24] The quality of photon antibunching is therefore directly related to energy transfer between both chromophores. We measured the statistics of fluorescence photons of a single 7 bp-distance sample by splitting the detection path onto two detectors which yield the coincidence counts in dependence of the lag time between the two detectors. Figure 3a plots the coincidence counts for a single 7 bp-distance sample acquired with laser pulses separated by 25 ns. The ratio of the magnitude of the central peak at lag time equals zero, $N_C$, to that of the lateral peaks, $N_L$, provides a measure for the degree of photon antibunching. For two completely independent chromophores, a value of 0.5 for $N_C/N_L$ is expected.[27] The example shown in Figure 4a, displays an antibunching value, $N_C/N_L$, of ~0.1, which translates to almost 100 % singlet-singlet-annihilation after considering the signal/background level.[28] $N_C/N_L$ was determined for 56 single 7 bp-distance samples plotted in a histogram in Figure 4b (black bars). The histogram shows a narrow distribution between 0 and 0.2. In comparison, the 68 single 20 bp-distance samples have mainly antibunching values around 0.5 (grey bars). In conclusion, singlet-singlet annihilation plays a significant role in the 7 bp-distance sample.

Long-lived dark states can be deliberately induced by simply removing oxygen, which leads to either long-lived triplet states or depending on the environment to long-lived radical states by a subsequent electron transfer. This is evidenced by strong blinking behavior of the dyes, which is shown in Figure 4c for a DNA origami structure bearing one ATTO647N dye under oxygen removal *without* stabilizing agents.[20] For such a transient, the average time is extracted for which the molecule is in a fluorescent state, $\langle\tau_{on}\rangle$, or in a dark state, $\langle\tau_{off}\rangle$ (longer transients are provided in the supporting information in Figure S4). The blue dashed line indicates the threshold, which was used to differentiate between a fluorescent or a dark state. $\tau_{on}$ ($\tau_{off}$) was determined for each state by counting the number of consecutive time bins for which the intensity was above (below) the threshold. For a single ATTO647N dye an average $\langle\tau_{on}\rangle = 0.99\ ms$ and $\langle\tau_{off}\rangle = 22.3\ ms$ were measured. For ATTO647N dyes at 20 bp-distance, $\langle\tau_{off}\rangle$ is approximately halved to 13.5 ms and the fluorescence intensity of the bursts is similar to that of a single ATTO647N due to independent blinking of both dyes (see Figure 4d). However, the 7 bp-distance sample shows an unexpected blinking behavior, because here the dark state of one dye is influencing the fluorescence properties of the neighboring dye. Figure 4e and 4f show two representative transients of the 7 bp-distance sample. 95 % of the transients can be sorted into these two types of blinking. In the first type (Figure 4e), two fluorescence intensity levels are clearly distinguishable: one at ~200-300 kHz (marked red) with a fluorescence lifetime of 4 ns (right panel, red curve); and a second with up to 60 kHz count rate (marked black) with a bi-exponential fluorescence lifetime decay (right panel, black curve) resulting in a long component with 4 ns and a short component of 1.1 ns. Further, $\langle\tau_{off}\rangle$ is strongly reduced to 3.33 ms. This behavior is explained by strong singlet-dark-state-annihilation (SDA), which also impacts the lifetime of the dark state itself. The transient in Figure 4e exhibits an example in which the energy transfer of one dye in its singlet manifold to the neighboring dye in its dark state is not 100 %, hence the remaining fluorescence during the dark state periods with a quenched fluorescence lifetime of 1.1 ns. The remaining 4 ns contribution stems from the small on-time periods, in which both dyes are in the singlet manifold. Additionally, this energy transfer leads to higher excited states of the dark state, for example higher triplet states, $T_n$, or excited radical states. Such higher excited dark states are more reactive, which can lead to a faster recovery to the ground state, $S_0$.[29,30] The transient in Figure 4f exhibits an example in which the energy transfer is almost 100 %, because no remaining fluorescence above the background was detected during the off-time periods. For this reason, the off-time periods are even more reduced to $\langle\tau_{off}\rangle = 1.25\ ms$, due to more efficient excitation and recovery of the dark state. We note that $\langle\tau_{on}\rangle$ is very similar for all cases with values below 1 ms, indicating that the time the dyes spend in the singlet manifold is poorly effected by energy transfer between them. However, energy transfer between the dyes in excited states must be considered and has two effects: (i) the negative effect is that the dark state of one dye is capable of quenching the fluorescence of dyes nearby; and (ii) the positive effect is that the dark state lifetime is reduced significantly leading to an higher brightness of the sample. As a consequence, dark states are detrimental especially in multi-chromophoric systems for achieving the highest brightness and can lead to saturation already at moderate excitation powers.[31] Singlet-singlet annihilation would only enhance saturation at high excitation powers.[32] Other transient dark states can fortunately be depopulated efficiently using the reducing and oxidizing system[20] so that a linear fluorescence response of a multichromophoric DNA origami structure can be obtained over more than an order of magnitude excitation power range and up to a count rate of 1 MHz for fluorescence detection (see Figure S5).

In summary, DNA origami are exciting scaffolds for dense dye arrangements to obtain bright point light sources. We here elaborated the rules to optimally place dyes in bright DNA origami constructs. The distance between two ATTO647N dyes was altered with single nucleotide step size. We revealed five relevant levels of interaction using single-molecule spectroscopy. Small distances show halved fluorescence lifetime and a tenfold decrease in intensity due to dynamic quenching and static quenching by H-type dimer formation. This contact related quenching is avoided when the dyes are separated by 7 bps when the spacer stems from the neighboring helix. With dye labels in the middle of the sequence our data indicate that 5 bp separation would be sufficient to avoid physical contact between the dyes. For the larger distances (e.g. 7 bps), weak coupling effects between dyes including singlet-singlet annihilation, singlet-triplet annihilation and singlet-radical-state annihilation have to be taken into account. Depending on the properties of the dyes used, the energy transfer between dyes can have beneficial or detrimental aspects on the fluorescence e.g. by leading to saturation or by opening new photophysical pathways for dark-state depopulation. The data shown also highlights the importance of avoiding dark state formation by using stabilizing agents.[20] This knowledge will pave the way to create small DNA origami nanobeads with unprecedented brightness density by preserving the photo physical properties of the dyes used. Based on the finding of a minimal spacing of 5 nucleotides between two dyes, we estimate that more than 1000 dyes could be placed within one DNA origami structure as used here without substantial perturbation of the photophysical properties. Despite interesting applications,[33] a detailed single-molecule study of fluorescence interactions of identical dyes had been lacking and will help in the design and interpretation of many biophysical single-molecule experiments.

ASSOCIATED CONTENT

**Supporting Information**. Details of the confocal fluorescence microscope, DNA origami folding, surface preparation and immobilization, absorption spectra and data analysis is given in the Supporting Information.

AUTHOR INFORMATION

**Corresponding Author**

*Philip.Tinnefeld@cup.lmu.de

*Jan.Vogelsang@cup.lmu.de

**Author Contributions**

The manuscript was written through contributions of all authors. All authors have given approval to the final version of the manuscript.

ACKNOWLEDGMENT

This work was funded by the DFG excellence clusters NIM (Nanosystems Initiative Munich), CIPSM (Center for Integrated Protein Science Munich) and e-conversion and by the European Union's Horizon 2020 research and innovation programme under grant agreement No 737089 (Chipscope).

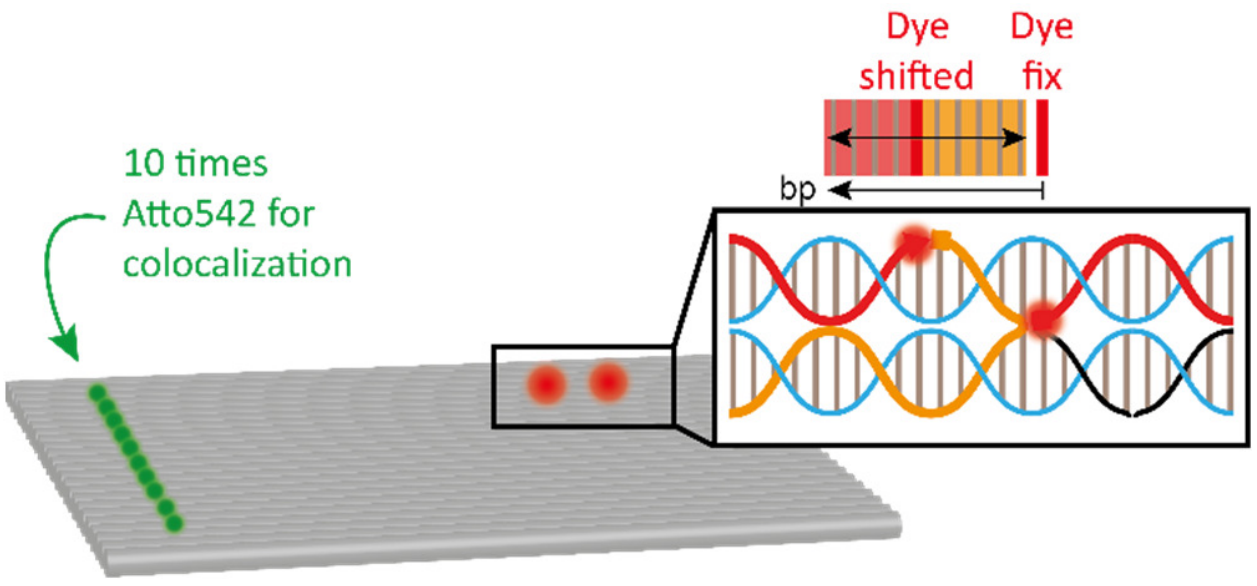

**Figure 1.** Model of the rectangular DNA origami labeled with up to ten green ATTO542 dyes for DNA nanostructure identification and two ATTO647N dyes for dye-dye-interaction studies. The spacer oligo (orange) hybridizes with the scaffold (light blue) to separate the dyes (red glowing dots) by a crossover from the neighboring helix. For every distance a different dye labeled oligo was used together with an adjusted spacer length. The right oligo stays the same for every experiment and is labeled at the 5' end. Other oligonucleotides are black. The magnified view shows a 6bp distance between the ATTO647N dyes.

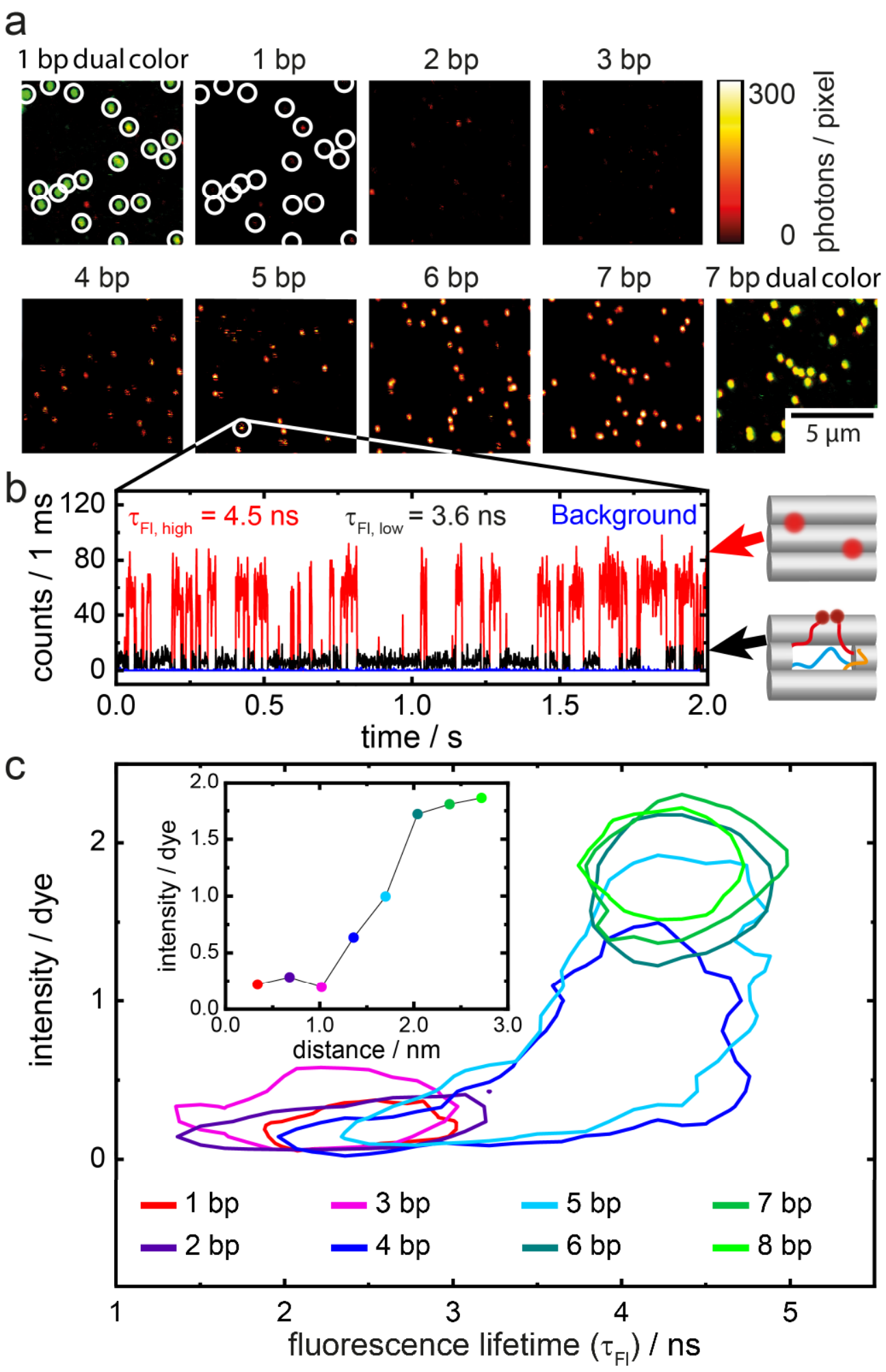

**Figure 2.** a) Confocal fluorescence microscopy images of DNA origami nanostructures under oxygen removal and ROXS using trolox/troloxquinone.[20, 34] The first image shows a false-color overlay image of the green reference channel and the red fluorescence from the 1bp-distance sample. The following images show the red spectral channel with increasing distance between the ATTO647N molecules. The last image is again an overlay of green and red channel to demonstrate perfect co-localization. Each pixel is 50 × 50 nm with an integration time of 1 ms for each color. b) Representative transient of a blinking 5bp-distance sample. High intensity corresponds to a closed stem and low intensity states correspond to an open stem. Both states are depicted schematically on the right in panel (b). (c) Sample brightness normalized to the average brightness

of one dye vs. fluorescence lifetime for the 1 bp- to 8 bp-distance samples. Lines surround 90 % of the overall population. The inset shows the mean intensity of each distribution versus the distance between both dyes in nanometers, assuming that 1bp corresponds to ~ 0.34 nm.

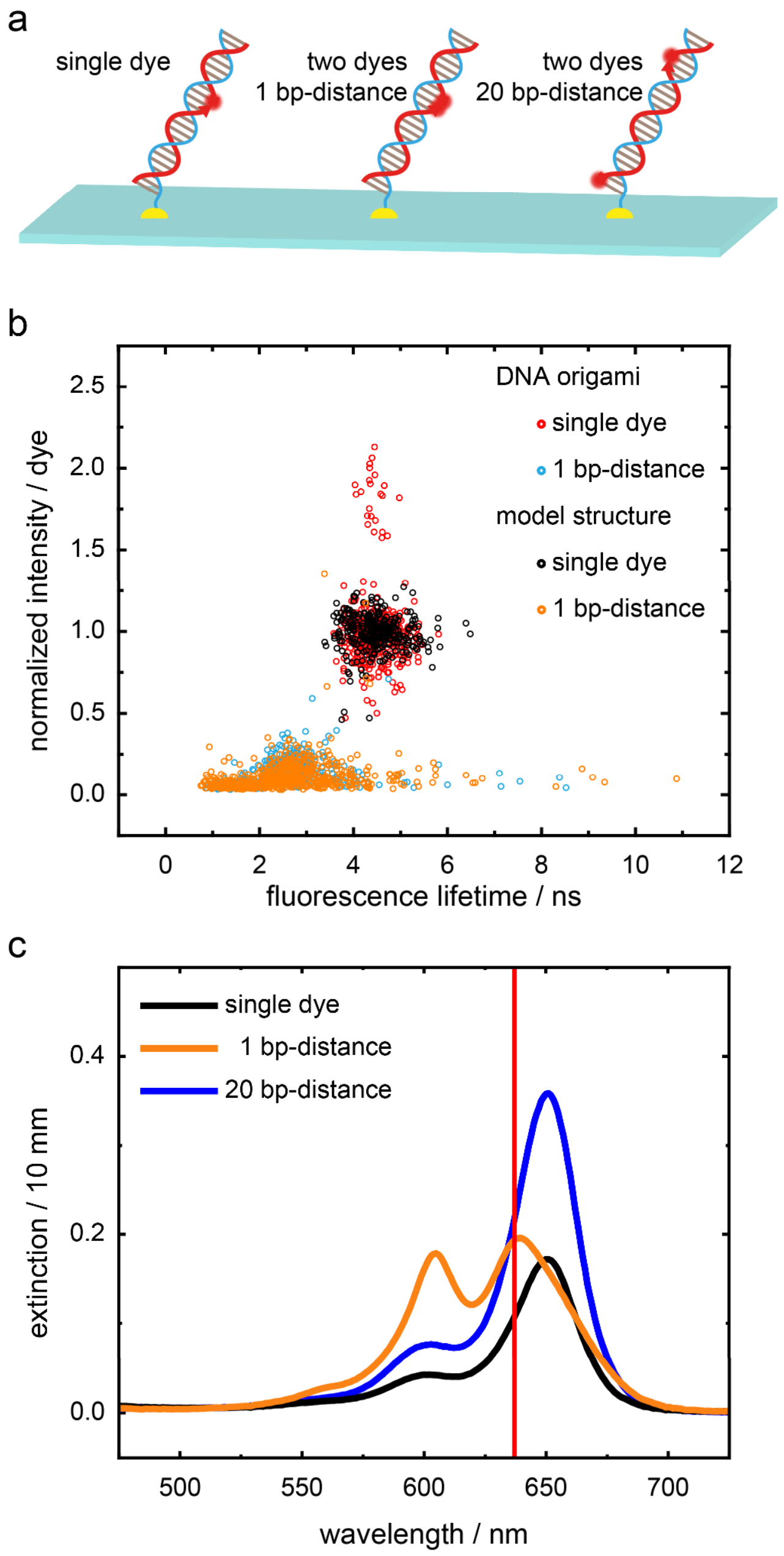

**Figure 3.** a) Schematic overview of the immobilized model structures used to compare with the DNA Origami structure and to measure the absorption spectrum of the dimer. b) Scatterplot of fluorescence lifetime and fluorescence intensity. DNA origami and model structure are compared and show no difference at the single molecule level for the single dye samples and the two dyes at 1 bp-distance samples. Spot finding was performed only on the red detection channel. A small

population of dimers is observed for the 1 bp-distance DNA origami, because DNA origami tend to blunt-end stack. c) Absorption spectra of the 1 bp-distance model (orange), 20 bp-distance model (blue) compared to a single Atto647N dye attached to dsDNA (black curve). All concentrations are set to 4.5 µM. The red line indicates the laser excitation wavelength of 637 nm at which all DNA origami samples are excited. We note that the extinction is very similar for the 1 bp- and 20 bp-distance sample.

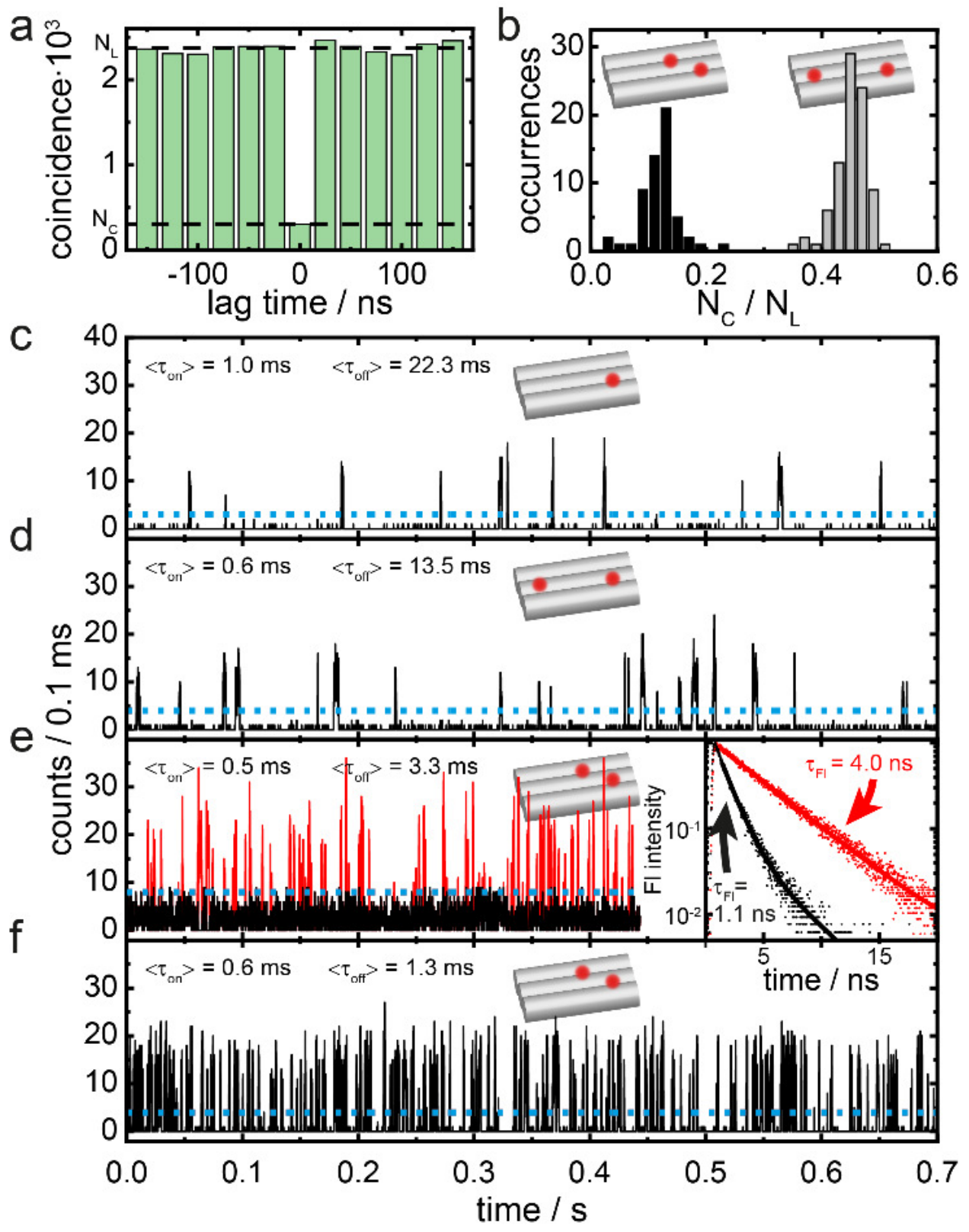

**Figure 4.** a) Photon antibunching from a 7-bp distance sample under oxygen removal and ROXS. The sample was excited by laser pulses (40 MHz repetition rate). The photon statistics in emission are shown in terms of coincidence counts of two photodetectors per one pulse delay in the emission pathway. The ratio between the center peak, $N_C$, and lateral peaks, $N_L$, are stated by dashed lines, respectively. b) Histograms of $N_C/N_L$ values for the 7 bp-distance (black bars, 56 measured spots) and 20bp-distance (grey bars, 68 measured spots) sample. c-f) Representative transients of a DNA-origami sample bearing only one ATTO647N dye (c), the 20bp-distance sample (d) and two examples of the 7 bp-distance sample (e, f) under oxygen removal without stabilizing agents. The average on- and off-times, $\tau_{on}$ and $\tau_{off}$, for which the molecule is in a fluorescent on-state or a non-fluorescent off-state are given for each transient. The blue dashed line is the threshold to separate the on- and off-states. We note that in panel (e) all intensity values corresponding to the (un)quenched state are color coded (red)black. From the corresponding photons the fluorescence decays and fluorescence lifetimes, $\tau_{Fl}$, are given for the bright state (red curve) and dark state (black curve).

TOC figure:

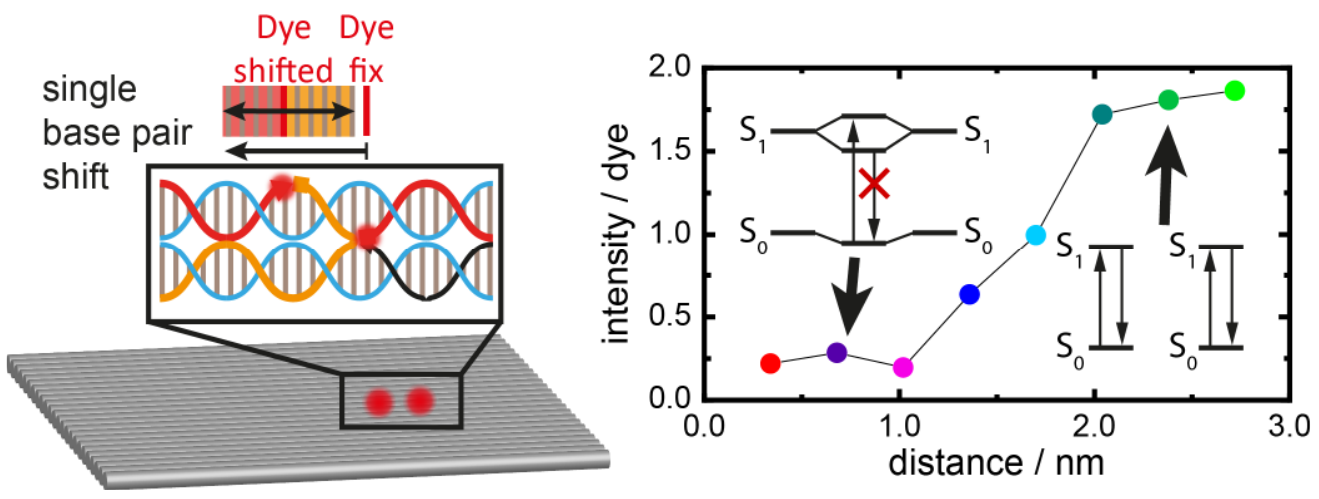

SI

# Interchromophoric Interactions Determine the Maximum Brightness Density in DNA Origami Structures

Tim Schröder, Max B. Scheible, Florian Steiner, Jan Vogelsang*, Philip Tinnefeld*

Department Chemie and Center for NanoScience, Ludwig-Maximilians-Universitaet Muenchen,

Butenandtstr. 5-13 Haus E, 81377 Muenchen, Germany.

## 1. Materials and Methods

### Confocal setup

A home built confocal microscope based on an Olympus IX-71 inverted microscope was used. Two pulsed lasers (637 nm, 80 MHz, LDH-D-C-640; 532 nm, 80 MHz, LDH-P- FA-530B; both PicoQuant GmbH) were altered through an acousto optical tunable filter (AOTFnc-VIS, AA Opto Electronic). Circular polarized light was obtained by a linear polarizer (LPVISE100-A, Thorlabs GmbH) and a quarter-wave plate (AQWP05M- 600, Thorlabs GmbH). The light was focused by an oil-immersion objective (UPLSAPO100XO, NA 1.40, Olympus Deutschland GmbH) onto the sample. The sample was moved by a piezo stage (P-517.3CD, Physik Instrumente (PI) GmbH & Co. KG) controlled by a E-727.3CDA piezo controller (Physik Instrumente (PI) GmbH & Co. KG). The emission was separated from the excitation beam by a dichroic beam splitter (z532/633, AHF analysentechnik AG) and focused onto a 50 µm pinhole (Thorlabs GmbH). The emission light was split by a dichroic beam splitter (640DCXR, AHF analysentechnik AG) into a green (Brightline HC582/75, AHF analysentechnik AG; RazorEdge LP 532, Laser 2000 GmbH) and red (Shortpass 750, AHF analysentechnik AG; RazorEdge LP 647, Laser 2000 GmbH) detection channel. Emission was focused onto avalanche photo diodes (SPCM-AQRH-14-TR, Excelitas Technologies GmbH & Co. KG) and signals were registered by a time-correlated single photon counting (TCSPC)-unit (HydraHarp400, PicoQuant GmbH). The setup was controlled by a homemade LabVIEW software or a commercial software package (SymPhoTime64, Picoquant GmbH). For cross correlation experiments the dichroic beam splitter in the detection path was substituted by a non-polarizing 50:50 beam splitter cube (CCM1-BS013/M, Thorlabs GmbH).

### DNA origami sample preparation

The flat rectangular DNA origami [1] was modified using caDNAno (version 0.2.2, design schematics in Fig. S6-S16). The 7249 nucleotide long scaffold was extracted from M13mp18 bacteriophages. All staple strands were purchased from Eurofins Genomics GmbH as well as the ATTO 647N modified oligos. The ATTO 542 modified oligos were purchased from biomers.net. For DNA origami folding oligos and scaffold from Table S1 were mixed for given final concentrations. As folding buffer (FB) 1x TAE with additional 12 mM $MgCl_2$ was used. For folding a nonlinear thermal annealing ramp over 16 hours was used [2]. After annealing the excess staples were removed by polyethylene glycol (PEG) precipitation [3]. The samples were mixed with an equal volume of PEG precipitation buffer (1x TAE, 15 % (w/v) PEG-8000, 500 mM NaCl, 12 mM $MgCl_2$) and centrifuged at 16 krcf for 30 min at 4°C. The pellet was suspended in 1x FB. Afterwards the DNA origami was externally labeled with ATTO 542 modified oligos. A three times excess respectively the extended staples was used and incubated overnight in a wet chamber. The DNA origami structures were purified via a gel electrophoresis. Therefor a 1.5 % agarose gel containing 0.5x TAE and 11 mM $MgCl_2$ was used at 70 V for 2 hours in a gel box cooled in an ice water bath. The gel was not stained to avoid unwanted staining reagent-dye interaction. DNA origami structures could be seen on a blue illuminated table due to the numerous ATTO 542 dyes. The target band was cut out and the DNA origami structures were recovered from the gel. The samples were stored at -26 °C until further use.

**Folding Table**

Final concentration for DNA origami folding are given in Table S1. Meaning of the reagents is described below:

**Table S1.** Folding reagents with final concentrations.

| **Reagent** | **Final concentration / nM** |
|---|---|
| scaffold | 30 |
| core staples | 240 |
| biotin staples | 300 |
| extended staples | 240 |
| dye used in every DNA origami | 450 |
| refill for 10bp | 240 |
| dye with different distance | 450 |

scaffold: single stranded viral DNA from M13mp18.

core staples: Contains every unmodified staples of the rectangular DNA origami. The wild structure is given in reference [1].

biotin staples: Six Biotin modified staples. Modifications are placed at the 5’ end.

extended staples: Ten staples extended at the 3' end for external labeling. The extended sequence is: 5' TTTTCCTCTACCACCTACATCAC 3'

dye used in every DNA origami: Oligo labeled at the 3' end with ATTO647N. This oligo is used in every DNA Origami.

refill for 10bp: For the 10bp sample the DNA origami was slightly modified to granite a stable incorporation of the 10bp oligo. This oligo is missing in the 10bp sample. For a 10bp sample replace it with $H_2O$.

dye with different distances: This stock contains the oligo labeled at the 5' end with its corresponding spacer.

The listed reagents were mixed and the folding buffer (FB) was added to 1xFB concentration.

### Surface preparation and immobilization

Measurements were performed in LabTek™ chamber slides (Thermo Fisher Scientific Inc.) which were cleaned two times for 20 minutes with 0.1 M hydrofluoric acid (AppliChem GmbH) and washed afterwards three times with ultrapure water. The glass surface was coated with biotin labeled bovine serum albumin (BSA) (Sigma-Aldrich Chemie GmbH). The DNA origami structure was immobilized through NeutrAvidin (Sigma-Aldrich Chemie GmbH) and six biotin labeled oligos to the BSA coated surface. dsDNA model structures were used likewise.

### Single molecule measurements

Surface scans were performed after DNA origami structure immobilization. An oxidizing and reducing buffer system (1x TAE, 12 mM $MgCl_2$, 2 mM Trolox/Troloxquinone, 1 % (w/v) D-(+)-Glycose) [4] was used in combination with an oxygen scavenging system (1 mg mL−1 glucose oxidase, 0.4 % (v/v) catalase (50 µg mL−1), 30 % glycerol, 12.5 mM KCl in 50 mM TRIS) to suppress blinking and photo bleaching. The oxygen scavenging system was added to the oxidation and reducing buffer at a concentration of 10 % (v/v) in the LabTek™. For surface scans a 10 × 20 $\mu m^2$ area size was used with a pixel size of 50 × 50 nm and alternating laser excitation. The integration time was 2 ms (1 ms for each color) and the laser power was adjusted to 9 µW at 639 nm and 1 µW at 532 nm.

Blinking kinetics under oxygen depletion without ROXS were performed by using glucose oxidase and catalase as described above. The measuring buffer was a 1xTEA buffer with 12 mM $MgCl_2$ and 1 % (w/v) D-(+)-Glycose.

### Absorption spectrum

For absorption spectra a higher concentration and amount of mass was needed than DNA origami structure folding could provide. Therefor a model structure was designed. Oligos, that are used in the DNA origami structure were hybridized to a complementary sequence (1bp: 5'- Biotin TTAATGAAACTTGATTCTGTCGCTACTGATTACGGTGCTGCTATCGATGGTTTCTGA, 20bp: 5'- Biotin TTAATGAAACTTGATTCTGTCGCTACTGATTACGGTGCTGCTATCGTGGTTTCTGAGGG TGGTGGCTCTTCAAGGCC). The spacer was the same as in the DNA origami structure. For hybridization equal amounts of substance were mixed and the solution was adjusted to the 1x FB concentration of TAE and $MgCl_2$. The solution was heated to 70 °C for 5 min and subsequently cooled down at a linear ramp to 25 °C by 1 K per minute. The absorption measurements were performed using a 10 mm path length cuvette (UVette, Eppendorf AG) and an Evolution 201 spectrometer (ThermoFisher Scientific Inc.) with 1 nm resolution and 1 s point integration time.

**Data analysis**

Each scan image has a 10 × 20 µm² size with a pixel size of 50 × 50 nm². Each pixel has a total integration time of 2 ms (1 ms per color). We use a home-build LabVIEW software with a spot finding algorithm to analyze the scans. DNA origami structures were marked with up to ten ATTO 542 dyes. Therefor the spot finding algorithm uses the green excitation green emission channel to find spots.

To define a spot, we used three different filters. The first one discriminates the pixels that we take into account. If a pixel has less or equal than 10 photons the algorithm does not take this pixel into account. The second filter discriminates by spot size. If an area of neighboring pixels, that were taken into account, is between 10 and 60 pixels we used them for further analysis. This is the expected area size of our PSFs. If an area is smaller, it is probably due to scattering dirt. A bigger area refers to two overlapping or close by DNA origami structures. The third parameter is the Heywood circular factor. Areas with a factor between 1.00 and 1.22 were taken into account. We use the last filter to get rid of PSFs which are cut in half because they are located at the edge of a scan. The remaining spots are analyzed. The program sums up the photons that are in range of a seven-pixel radius from the center of the spot for each channel. Red excitation, red emission channel was used to obtain the intensity per spot and fluorescence lifetime data of the ATTO 647N dyes.

For the dsDNA model a green dye was not attached to the dsDNA. The red channel was analyzed with following parameters. A pixel was taken into account when it had more than 5 photons and a PSF was recognized when it had more than 10 and less than 80 pixels. The Heywood circular factor was not used due to the blinking characteristic of the dye dimer that leads to rough shaped spots.

When analyzing cross correlation measurements, the first three million photons were used to calculate the coincidence histogram. This gives a lateral count of roughly 2000 coincidences with little shot noise. Signal to background ratio for one dye experiments was always above 170 and for two dyes above 340. For calculation the coincidence ratio the central peak was divided by the average of the six lateral peaks.

**2. AFM**

AFM imaging was performed on a NanoWizard® 3 ultra AFM (JPK Instruments AG) in solution using 1xFB. The DNA origami structures were immobilized on a freshly cleaved mica surface (Quality V1, Plano GmbH) by $Ni^{2+}$ ions which were incubated on the mica for 5 minutes with a 10 mM $NiCl_2$ solution. Afterwards the mica was washed three times with ultra-pure water and dried by dry air. The origami structures were incubated for 5 minutes by a 1 nM solution. Measurements were performed with a USC-F0.3-k0.3-10 cantilever from NanoWorld AG.

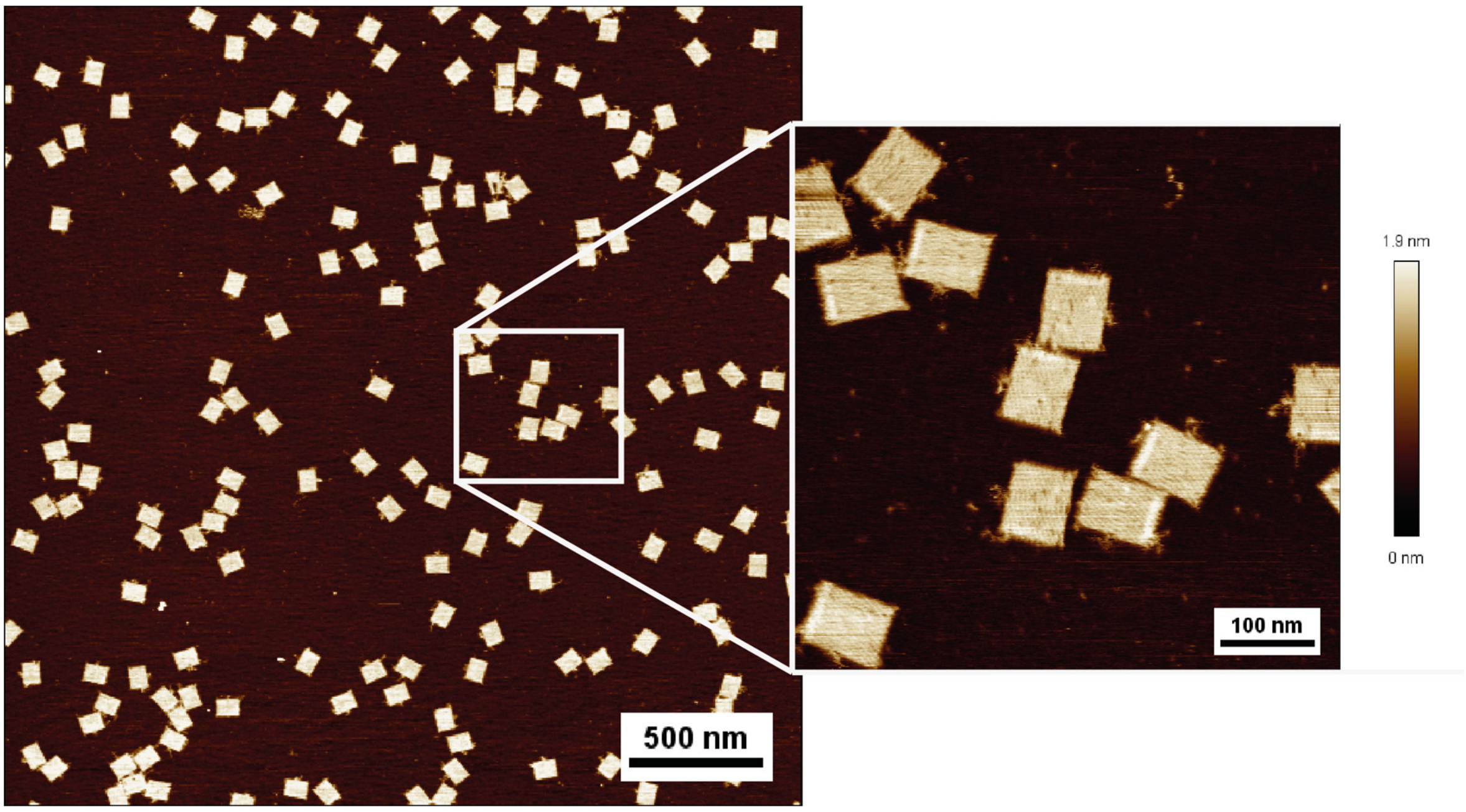

**Figure S1.** AFM image of a 1-bp distance sample. Magnification of the indicated area on the right. The bright lines on the DNA origamis indicate the external labeling due to an additional layer of dsDNA.

## 2. Blinking kinetics in the 4 bp- to 6 bp-distance samples

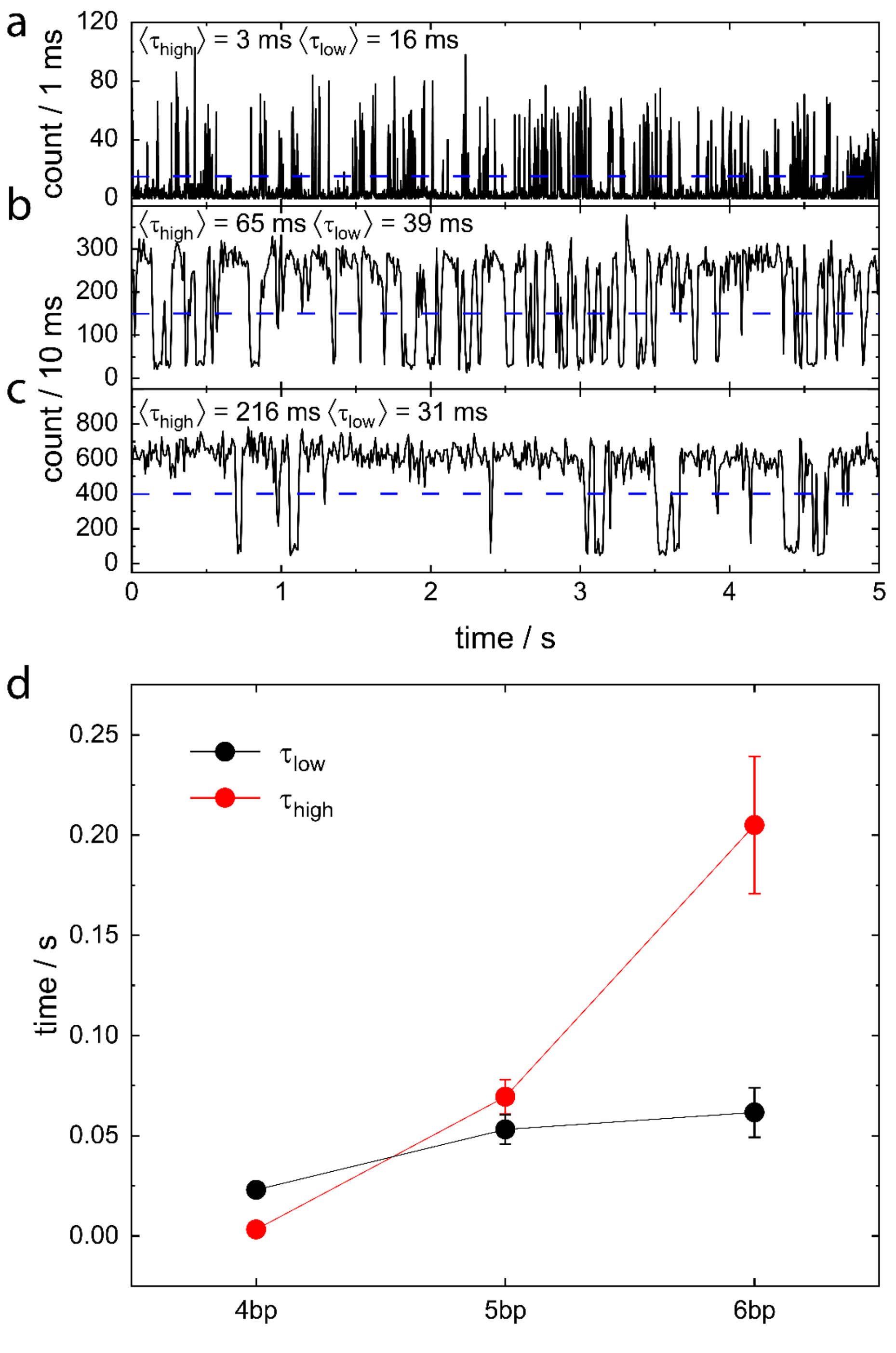

**Figure S2.** Representative transients from a) 4bp- b) 5bp- and c) 6bp-distance samples. Average dwell times for the high and low fluorescence state for each transient were extracted by using the threshold indicated as blue dashed lines. d) Mean dwell times are shown for each fluorescent state with standard error of the mean of 55, 47 and 32 DNA origami structures for the 4bp-, 5bp- and 6bp-distance samples, respectively. The lifetime of the high fluorescent state becomes longer with increasing spacer length, because additional nucleotides increase the hybridization energy of double stranded DNA. The lifetime of the low fluorescent state stays roughly the same, because the chance of hybridization is limited by the local concentration of the single stranded spacer sequence which is approximately the same for all samples. We note that the blinking kinetic is inhomogeneous throughout the samples due to different nano environments. Different nano environments mainly arises by magnesium ion which clip tow neighboring helices and can therefor stabilize the high fluorescent state.[5]

### 3. Quenching in the 1 bp- to 10 bp-distance samples without a spacer strand

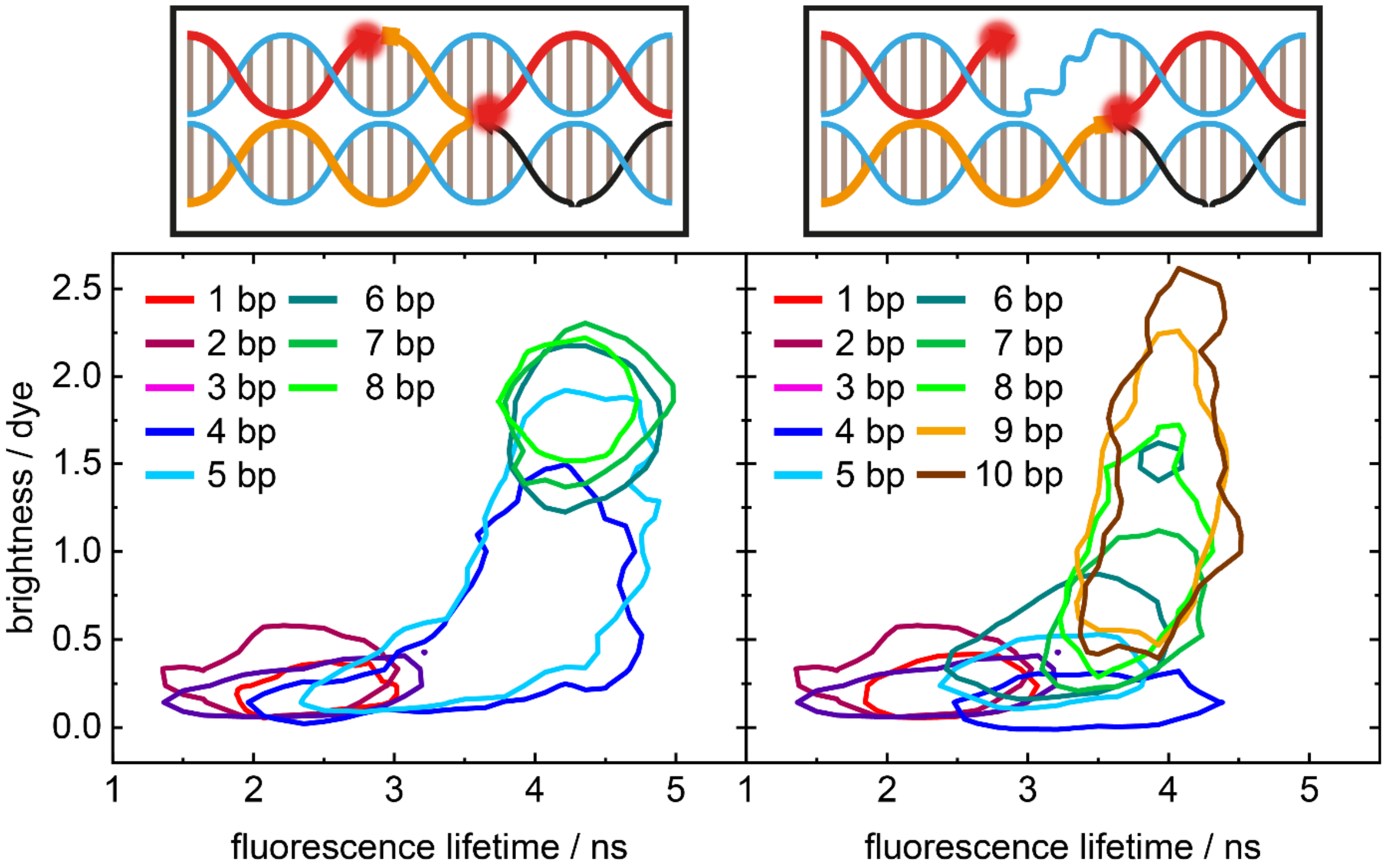

**Figure S3.** Sample brightness normalized to the average brightness of one dye vs. fluorescence lifetime for the 1 bp- to 10 bp-distance samples with (left panel) and without (right panel) a spacer strand. Lines surround 90 % of the overall population. No pure high intensity and long lifetime population is visible in the samples without a spacer strand.

## 4. Transients of the 7 bp-distance sample

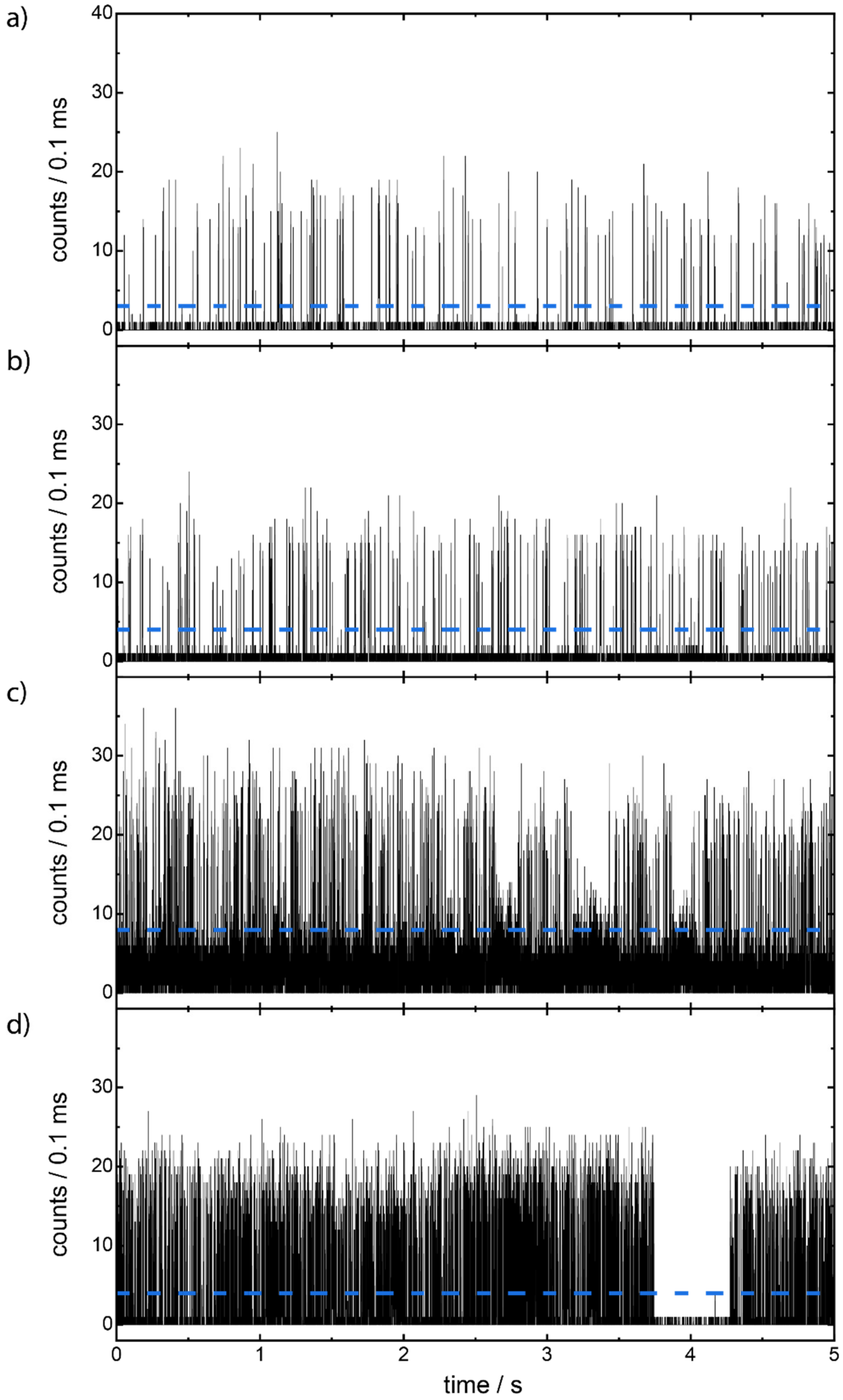

**Figure S4.** Extended transients from Figure 3 c)-f) with oxygen scavenging and without further stabilization. Blue lines indicate threshold for the determination of on- and off-times, respectively.

a) DNA origami sample with one ATTO647N dye, b) 20 bp-distance sample, c) 7 bp-distance sample with background fluorescence and d) without background fluorescence.

## 5. Fluorescence intensity dependence of the 7 bp-distance sample

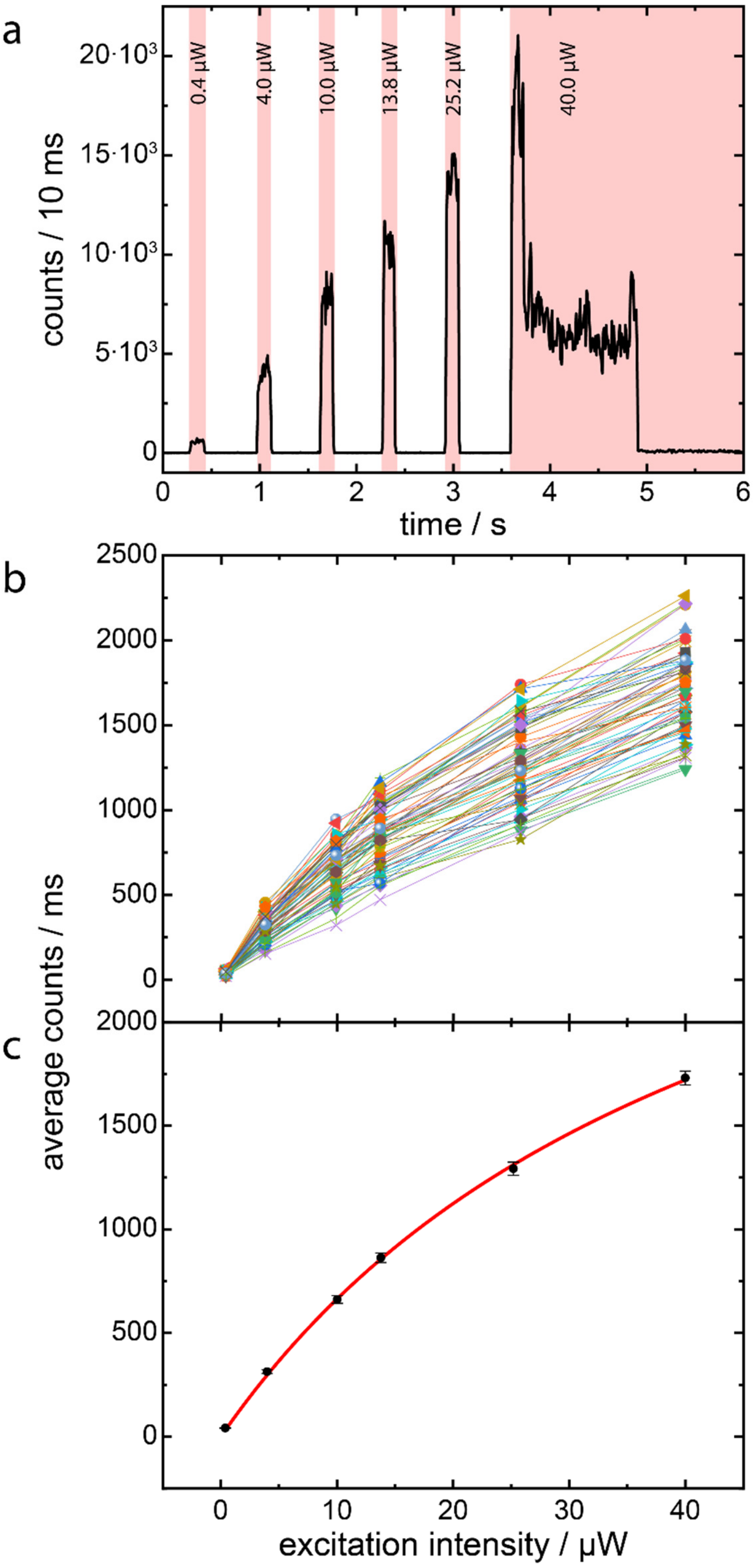

**Figure S5.** a) Fluorescence transient of a 7 bp-distance sample with increasing laser excitation intensities as stated in the panel. Red areas highlight laser excitation. Two bleaching steps at the end confirm the presence of two dyes. The count rate was extracted at each excitation intensity. b)

Count rate at the set excitation powers of 61 transients. c) Averaged count values obtained from the 61 transients shown in (b) with standard error of the mean. A saturation curve in red is fitted to the data points to obtain the maximum count rate = $3.6 \cdot 10^6$ Hz. The following model was used:[6] $N = \frac{N_{max} \cdot \frac{I_{ex}}{I_{sat}}}{1 + \frac{I_{ex}}{I_{sat}}}$ where N is the count rate, $N_{max}$ the maximum count rate, $I_{ex}$ the excitation intensity and $I_{sat}$ represents the excitation intensity where half of $N_{max}$ is reached. An almost linear fluorescence dependence is demonstrated up to an average count rate of ~ 1 MHz. An optical density filter of OD1 was placed in the detection path and corrected for to avoid saturation due to the dead time of the avalanche photo detector. The maximum count rate of $3.6 \cdot 10^6$ Hz translates to an emission rate ~$3.6 \cdot 10^7$ Hz under the assumption that the confocal microscope provides a detection efficiency of ~10 %. This count rate leads to a maximum off-time of ~ 28 ns, which can be well attributed to a diffusion-limited quenching process of the triplet state by the reducing and oxidizing system at 1 mM concentration.

## 6. DNA origamis

Two exemplary caDNAno designs are shown below. Previously used staples are colored in black. Green staples are labeled at the 5' end with biotin. Yellow marked staples are extended at the 3' end with following sequence: 5' TTTTCCTCTACCACCTACATCAC 3'. Red staples are labeled with ATTO647N either at the 3' or 5'. Blue staples are staples around the ATTO647N labeled staples to stabilize the structure.

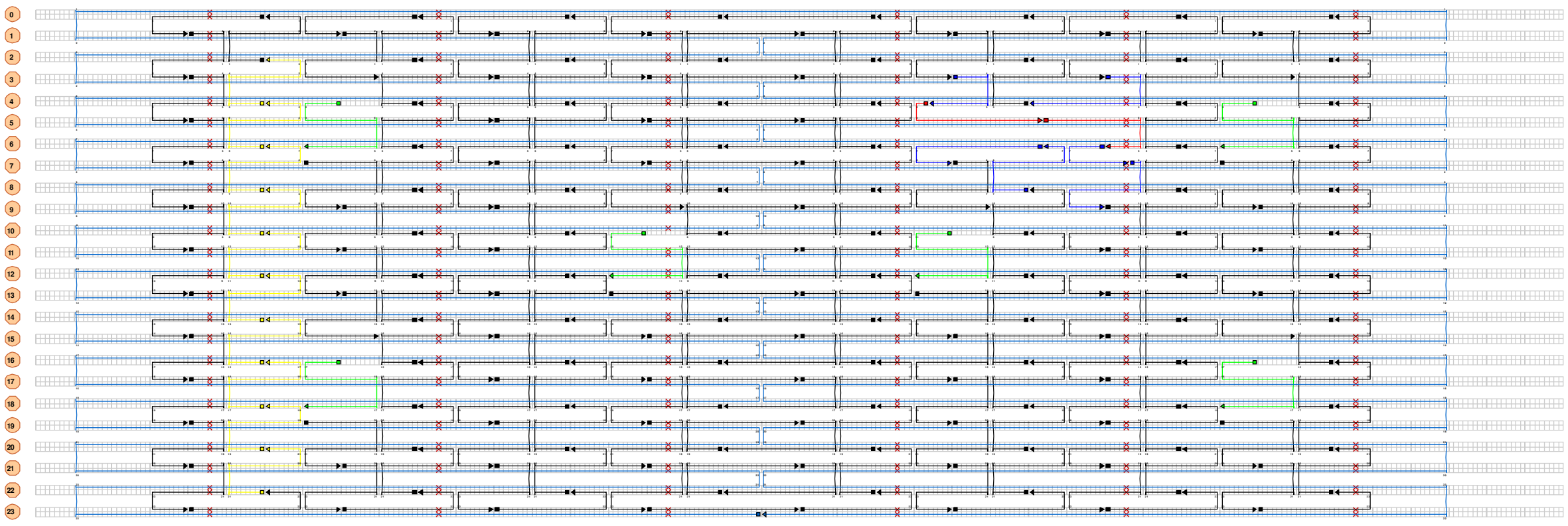

**Figure S6:** Design of the 1bp sample.

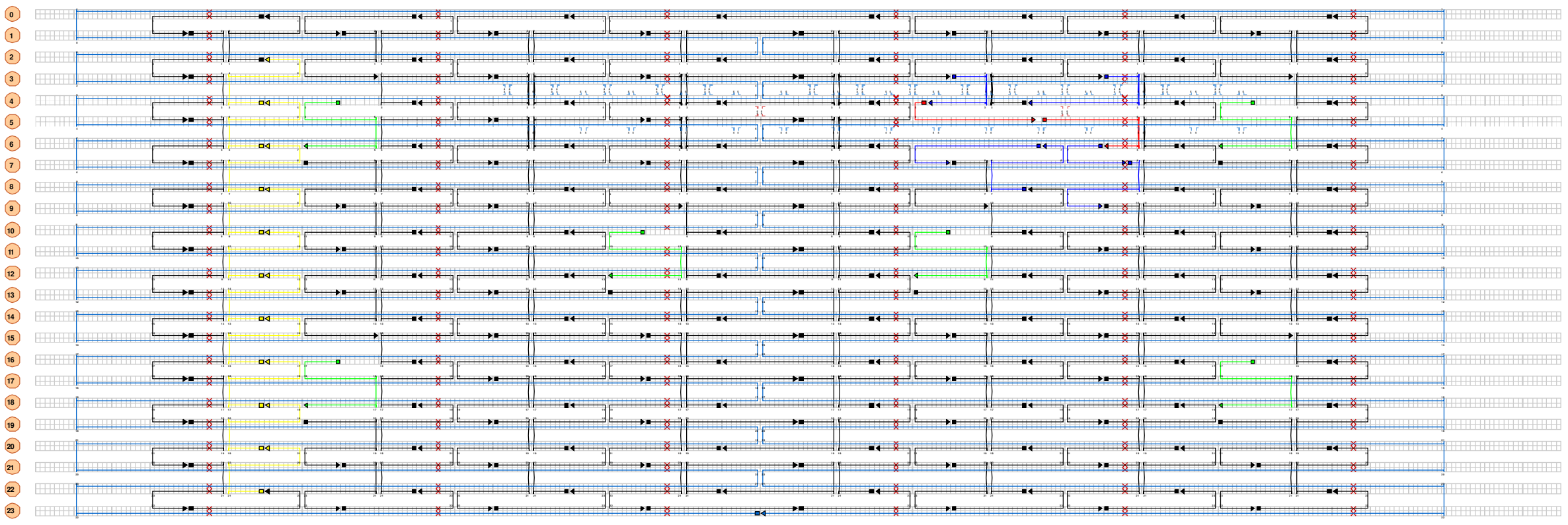

**Figure** S7: Design of the 2bp sample.

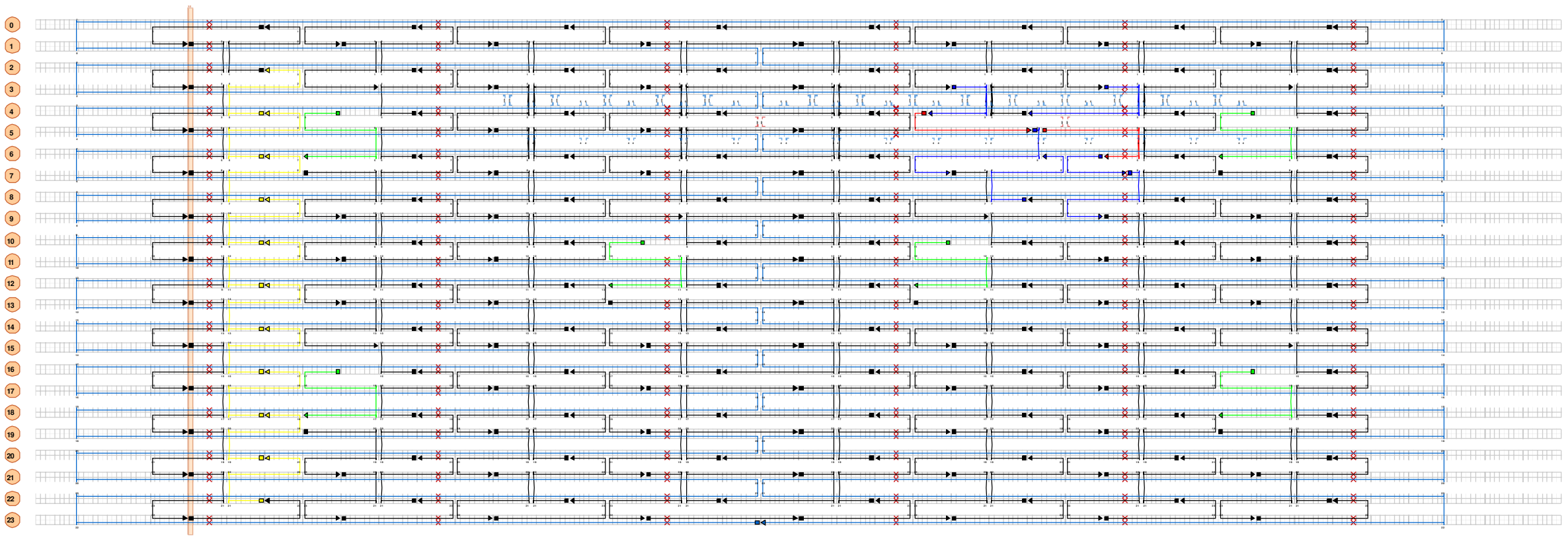

**Figure S8:** Design of the 3bp sample.

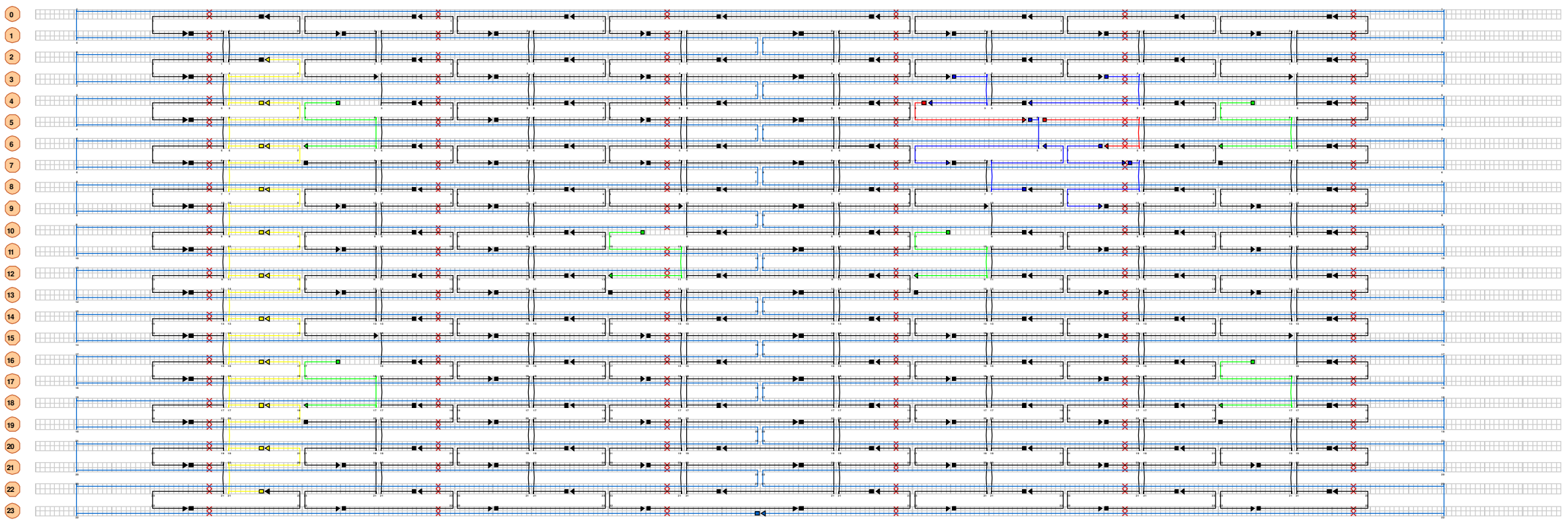

**Figure S9:** Design of the 4bp sample.

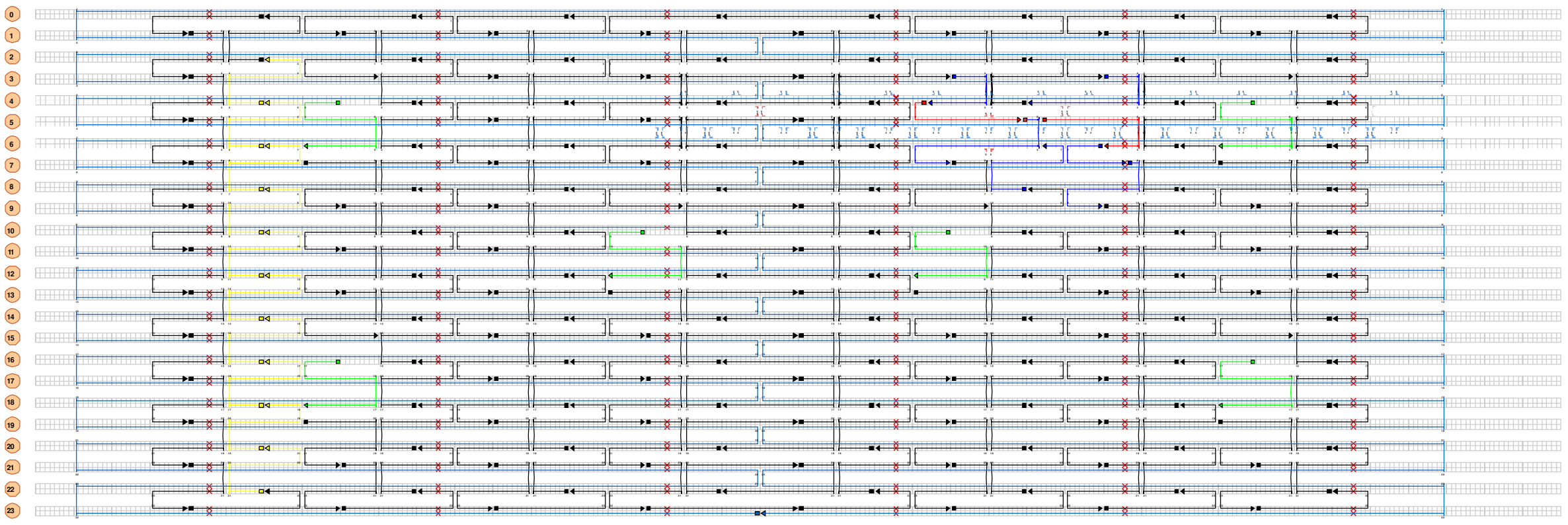

**Figure S10:** Design of the 5bp sample.

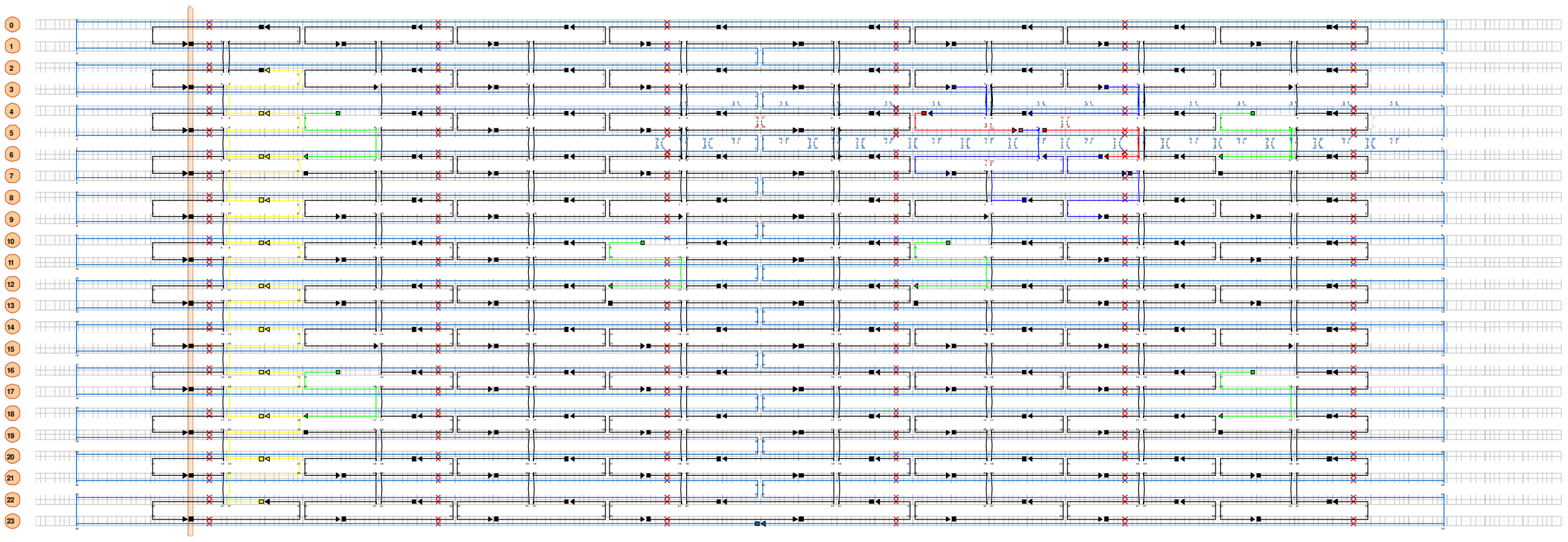

**Figure S11:** Design of the 6bp sample.

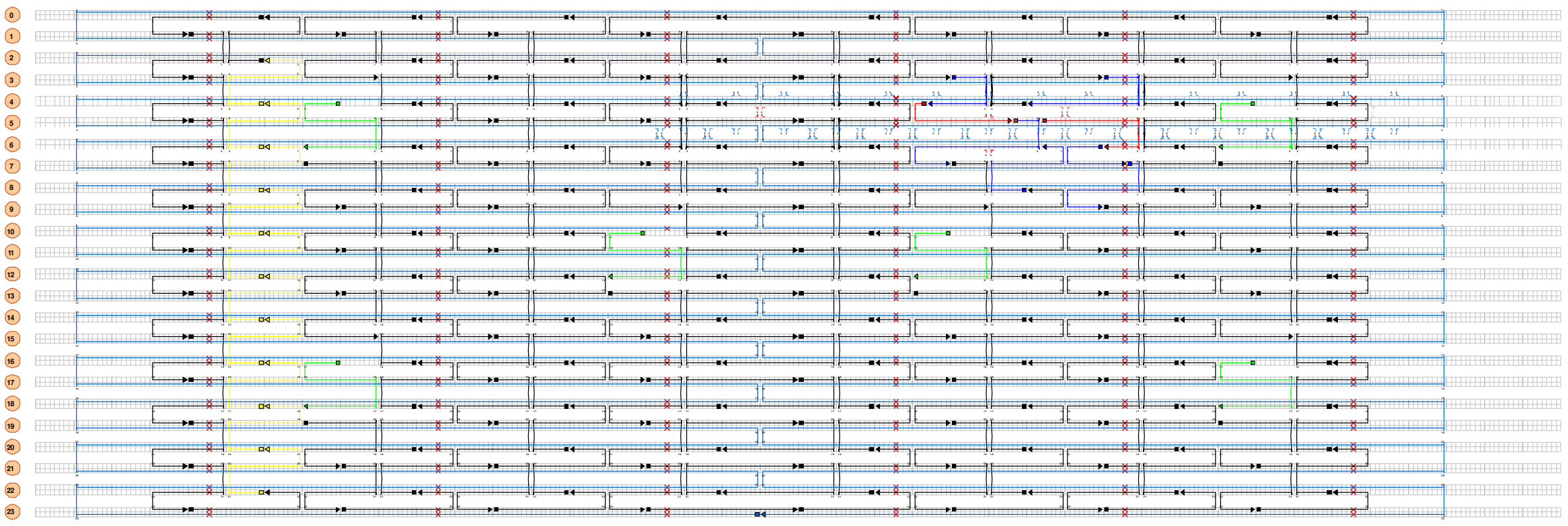

**Figure S12:** Design of the 7bp sample.

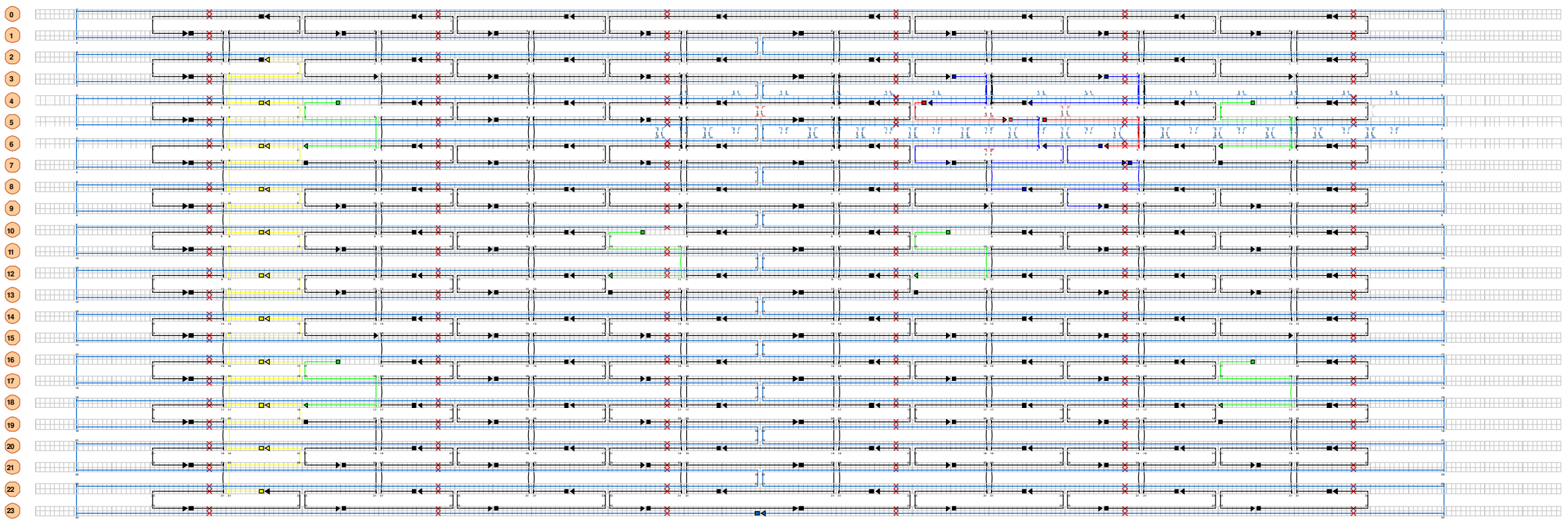

**Figure S13:** Design of the 8bp sample.

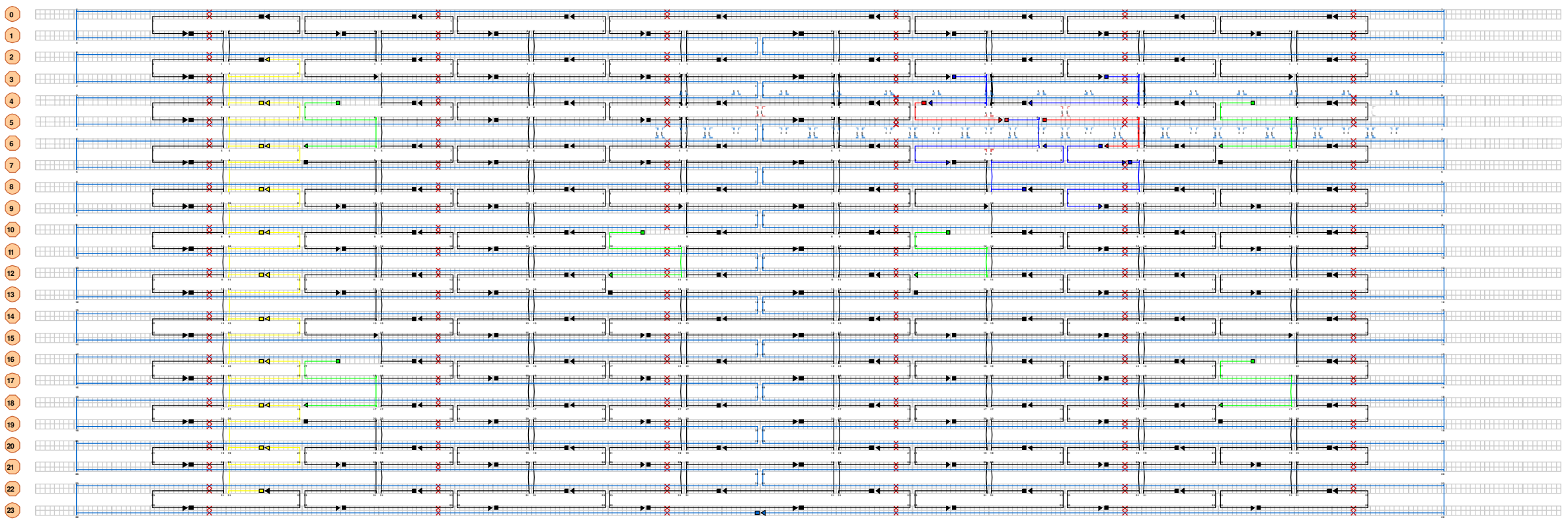

**Figure S14:** Design of the 9bp sample.

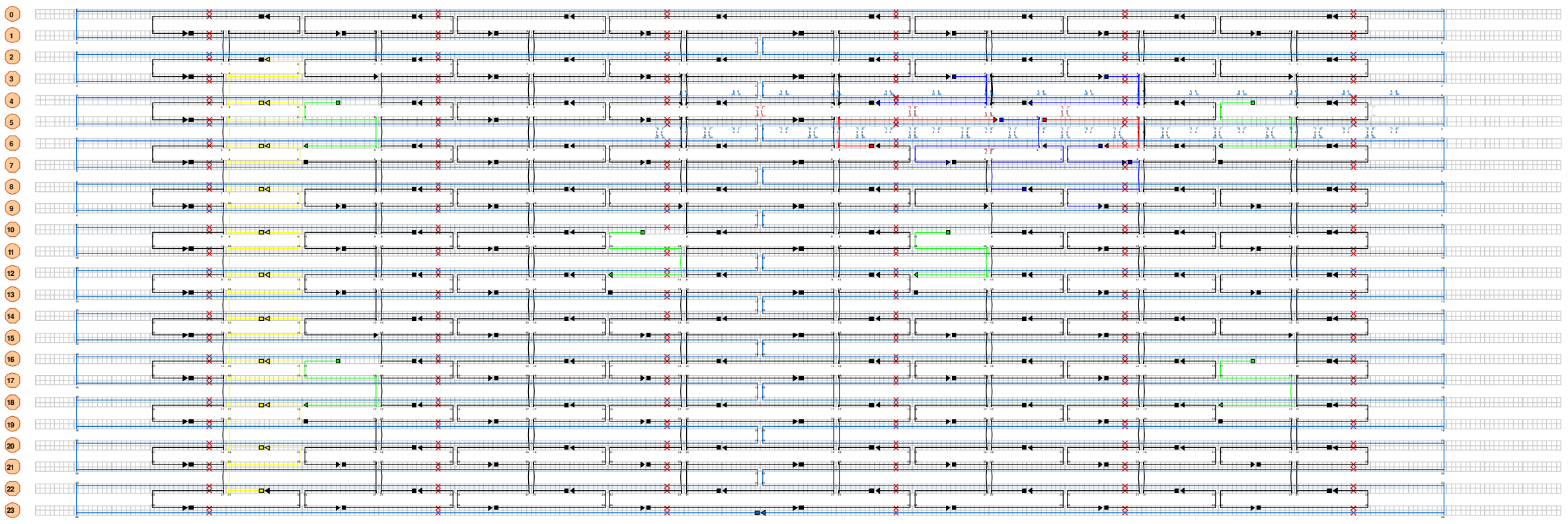

**Figure S15:** Design of the 10bp sample.

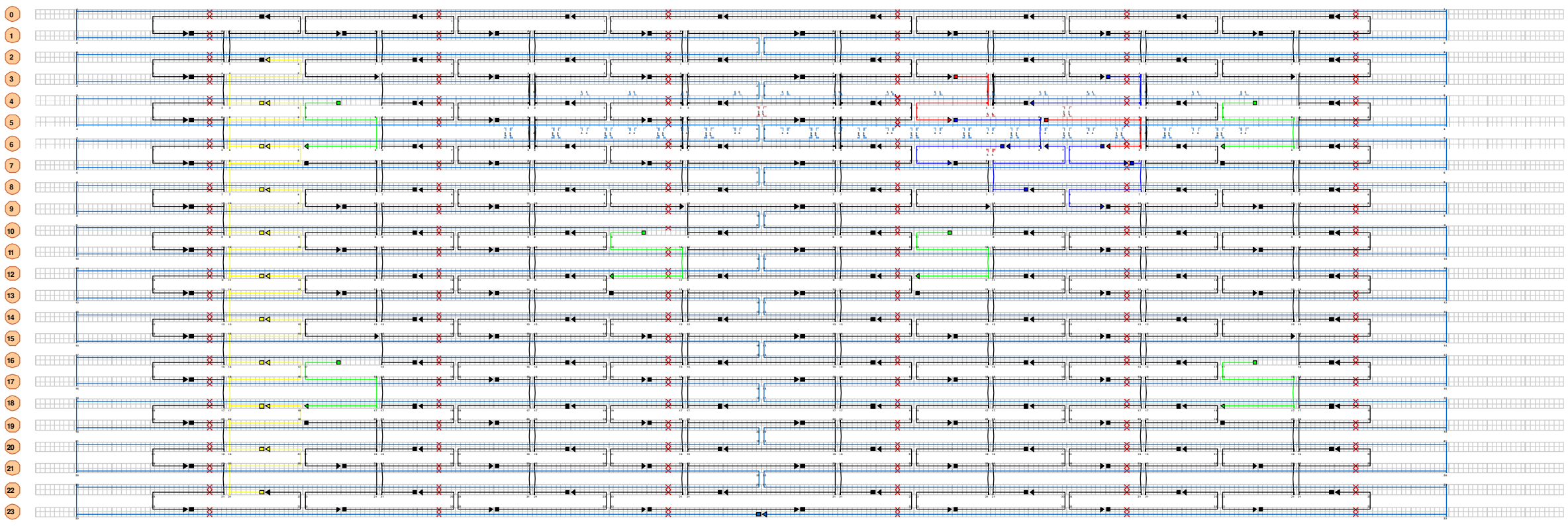

**Figure S16:** Design of the 20bp sample.

**Table S2:** Sequences of staples

| 5’ position | Sequence | Comment |
|---|---|---|
| 5[211] | AGTAGCGACAGAATCAAGTTTCATTAA | 5' labeled ATTO647N |
| 4[186] | TCAGAAACCATCGATAGCAGCACCGTAATC | 3' labeled ATTO64 1bp |
| 4[186] | TCAGAAACCATCGATAGCAGCACCGTAAT | 3' labeled ATTO64 2bp |
| 4[186] | TCAGAAACCATCGATAGCAGCACCGTAA | 3' labeled ATTO64 3bp |
| 4[186] | TCAGAAACCATCGATAGCAGCACCGTA | 3' labeled ATTO64 4bp |
| 4[186] | TCAGAAACCATCGATAGCAGCACCGT | 3' labeled ATTO64 5bp |
| 4[186] | TCAGAAACCATCGATAGCAGCACCG | 3' labeled ATTO64 6bp |
| 4[186] | TCAGAAACCATCGATAGCAGCACC | 3' labeled ATTO64 7bp |
| 4[186] | TCAGAAACCATCGATAGCAGCAC | 3' labeled ATTO64 8bp |
| 4[186] | TCAGAAACCATCGATAGCAGCA | 3' labeled ATTO64 9bp |
| 4[186] | CAGCAAAAGGAAACGTCACCAATGAAACCATCGATAGCAGC | 3' labeled ATTO64 10bp |
| 3[192] | GGCCTTGAAGAGCCACCACCCTCAGAAACCAT | 3' labeled ATTO64 20bp |
| 6[210] | CCGTCACCGACTTGAGCCATTTGGGAACGTAGAAA | spacer 1bp |
| 6[210] | TCCCGTCACCGACTTGAGCCATTTGGGAACGTAGAAA | spacer 3bp |
| 6[210] | ATCCCGTCACCGACTTGAGCCATTTGGGAACGTAGAAA | spacer 4bp |
| 6[210] | AATCCCGTCACCGACTTGAGCCATTTGGGAACGTAGAAA | spacer 5bp |
| 6[210] | TAATCCCGTCACCGACTTGAGCCATTTGGGAACGTAGAAA | spacer 6bp |
| 6[210] | GTAATCCCGTCACCGACTTGAGCCATTTGGGAACGTAGAAA | spacer 7bp |
| 6[210] | CGTAATCCCGTCACCGACTTGAGCCATTTGGGAACGTAGAAA | spacer 8bp |
| 6[210] | CCGTAATCCCGTCACCGACTTGAGCCATTTGGGAACGTAGAAA | spacer 9bp |
| 6[175] | ACCGTAATCCCGTCACCGACTTGAGCCATTTGGGAACGTAGAAA | spacer 10bp |
| 7[229] | CAAAGATAGCCGAACAAACCCTGAAC | stabilizer |
| 6[223] | AGGTGAATATAAAAGAAACG | stabilizer |
| 3[224] | TTAAAGCCAGAGCCGCCACCCTCAGAACCG | stabilizer |
| 8[207] | AAGGAAACATAAAGGTGGCAACATTATCA | stabilizer |
| 3[192] | GGCCTTGAAGAGCCACCACCC | stabilizer |
| 18[47] | CCAGGGTTGCCAGTTTGAGGGGACCCGTGGGA | for external labeling |
| 12[47] | TAAATCGGGATTCCCAATTCTGCGATATAATG | for external labeling |
| 22[47] | CTCCAACGCAGTGAGACGGGCAACCAGCTGCA | for external labeling |
| 16[47] | ACAAACGGAAAAGCCCCAAAAACACTGGAGCA | for external labeling |
| 8[47] | ATCCCCCTATACCACATTCAACTAGAAAAATC | for external labeling |
| 4[47] | GACCAACTAATGCCACTACGAAGGGGGTAGCA | for external labeling |
| 10[47] | CTGTAGCTTGACTATTATAGTCAGTTCATTGA | for external labeling |
| 14[47] | AACAAGAGGGATAAAAATTTTTAGCATAAAGC | for external labeling |
| 6[47] | TACGTTAAAGTAATCTTGACAAGAACCGAACT | for external labeling |
| 20[47] | TTAATGAACTAGAGGATCCCCGGGGGGTAACG | for external labeling |
| 10[191] | GAAACGATAGAAGGCTTATCCGGTCTCATCGAGAACAAGC | biotin |
| 10[127] | TAGAGAGTTATTTTCATTTGGGGATAGTAGTAGCATTA | biotin |
| 16[255] | GAGAAGAGATAACCTTGCTTCTGTTCGGGAGAAACAATAA | biotin |
| 4[255] | AGCCACCACTGTAGCGCGTTTTCAAGGGAGGGAAGGTAAA | biotin |
| 4[63] | ATAAGGGAACCGGATATTCATTACGTCAGGACGTTGGGAA | biotin |
| 16[63] | CGGATTCTGACGACAGTATCGGCCGCAAGGCGATTAAGTT | biotin |
| 6[175] | CAGCAAAAGGAAACGTCACCAATGAGCCGC | for 10bp not needed |
| 15[128] | TAAATCAAAATAATTCGCGTCTCGGAAACC | |
| 14[271] | TTAGTATCACAATAGATAAGTCCACGAGCA | |
| 17[224] | CATAAATCTTTGAATACCAAGTGTTAGAAC | |
| 8[175] | ATACCCAACAGTATGTTAGCAAATTAGAGC | |
| 19[248] | CGTAAAACAGAAATAAAAATCCTTTGCCCGAAAGATTAGA | |
| 5[96] | TCATTCAGATGCGATTTTAAGAACAGGCATAG | |
| 0[79] | ACAACTTTCAACAGTTTCAGCGGATGTATCGG | |
| 12[79] | AAATTAAGTTGACCATTAGATACTTTTGCG | |
| 6[111] | ATTACCTTTGAATAAGGCTTGCCCAAATCCGC | |
| 11[224] | GCGAACCTCCAAGAACGGGTATGACAATAA | |
| 16[111] | TGTAGCCATTAAAATTCGCATTAAATGCCGGA | |
| 13[120] | AAAGGCCGGAGACAGCTAGCTGATAAATTAATTTTTGT | |
| 16[271] | CTTAGATTTAAGGCGTTAAATAAAGCCTGT | |
| 11[96] | AATGGTCAACAGGCAAGGCAAAGAGTAATGTG | |
| 22[143] | TCGGCAAATCCTGTTTGATGGTGGACCCTCAA | |
| 5[128] | AACACCAAATTTCAACTTTAATCGTTTACC | |
| 20[271] | CTCGTATTAGAAATTGCGTAGATACAGTAC | |
| 15[224] | CCTAAATCAAAATCATAGGTCTAAACAGTA | |
| 21[96] | AGCAAGCGTAGGGTTGAGTGTTGTAGGGAGCC | |
| 1[224] | GTATAGCAAACAGTTAATGCCCAATCCTCA | |
| 2[143] | ATATTCGGAACCATCGCCCACGCAGAGAAGGA | |
| 4[79] | GCGCAGACAAGAGGCAAAAGAATCCCTCAG | |
| 13[184] | GACAAAAGGTAAAGTAATCGCCATATTTAACAAAACTTTT | |
| 6[239] | GAAATTATTGCCTTTAGCGTCAGACCGGAACC | |
| 14[111] | GAGGGTAGGATTCAAAAGGGTGAGACATCCAA | |
| 19[96] | CTGTGTGATTGCGTTGCGCTCACTAGAGTTGC | |
| 23[192] | ACCCTTCTGACCTGAAAGCGTAAGACGCTGAG | |
| 18[111] | TCTTCGCTGCACCGCTTCTGGTGCGGCCTTCC | |
| 4[111] | GACCTGCTCTTTGACCCCCAGCGAGGGAGTTA | |
| 18[239] | CCTGATTGCAATATATGTGAGTGATCAATAGT | |
| 21[224] | CTTTAGGGCCTGCAACAGTGCCAATACGTG | |
| 16[239] | GAATTTATTTAATGGTTTGAAATATTCTTACC | |
| 7[32] | TTTAGGACAAATGCTTTAAACAATCAGGTC | |
| 7[248] | GTTTATTTTGTCACAATCTTACCGAAGCCCTTTAATATCA | |
| 2[239] | GCCCGTATCCGGAATAGGTGTATCAGCCCAAT | |
| 12[271] | TGTAGAAATCAAGATTAGTTGCTCTTACCA | |
| 15[160] | ATCGCAAGTATGTAAATGCTGATGATAGGAAC | |
| 10[207] | ATCCCAATGAGAATTAACTGAACAGTTACCAG | |
| 10[271] | ACGCTAACACCCACAAGAATTGAAAATAGC | |
| 18[79] | GATGTGCTTCAGGAAGATCGCACAATGTGA | |
| 13[160] | GTAATAAGTTAGGCAGAGGCATTTATGATATT | |
| 20[207] | GCGGAACATCTGAATAATGGAAGGTACAAAAT | |
| 23[256] | CTTTAATGCGCGAACTGATAGCCCCACCAG | |
| 17[128] | AGGCAAAGGGAAGGGCGATCGGCAATTCCA | |
| 4[207] | CCACCCTCTATTCACAAACAAATACCTGCCTA | |
| 21[192] | TGAAAGGAGCAAATGAAAAATCTAGAGATAGA | |
| 7[160] | TTATTACGAAGAACTGGCATGATTGCGAGAGG | |
| 10[111] | TTGCTCCTTTCAAATATCGCGTTTGAGGGGGT | |
| 22[175] | ACCTTGCTTGGTCAGTTGGCAAAGAGCGGA | |
| 10[239] | GCCAGTTAGAGGGTAATTGAGCGCTTTAAGAA | |
| 1[64] | TTTATCAGGACAGCATCGGAACGACACCAACCTAAAACGA | |
| 14[143] | CAACCGTTTCAAATCACCATCAATTCGAGCCA | |
| 1[96] | AAACAGCTTTTTGCGGGATCGTCAACACTAAA | |
| 6[143] | GATGGTTTGAACGAGTAGTAAATTTACCATTA | |
| 14[239] | AGTATAAAGTTCAGCTAATGCAGATGTCTTTC | |
| 16[175] | TATAACTAACAAAGAACGCGAGAACGCCAA | |
| 12[175] | TTTTATTTAAGCAAATCAGATATTTTTTGT | |
| 2[175] | TATTAAGAAGCGGGGTTTTGCTCGTAGCAT | |
| 21[64] | GCCCTTCAGAGTCCACTATTAAAGGGTGCCGT | |
| 9[160] | AGAGAGAAAAAAATGAAAATAGCAAGCAAACT | |
| 12[207] | GTACCGCAATTCTAAGAACGCGAGTATTATTT | |
| 19[192] | ATTATACTAAGAAACCACCAGAAGTCAACAGT | |
| 0[175] | TCCACAGACAGCCCTCATAGTTAGCGTAACGA | |
| 1[128] | TGACAACTCGCTGAGGCTTGCATTATACCA | |
| 14[175] | CATGTAATAGAATATAAAGTACCAAGCCGT | |
| 7[56] | ATGCAGATACATAACGGGAATCGTCATAAATAAAGCAAAG | |
| 22[79] | TGGAACAACCGCCTGGCCCTGAGGCCCGCT | |
| 17[96] | GCTTTCCGATTACGCCAGCTGGCGGCTGTTTC | |
| 21[32] | TTTTCACTCAAAGGGCGAAAAACCATCACC | |
| 13[32] | AACGCAAAATCGATGAACGGTACCGGTTGA | |
| 6[271] | ACCGATTGTCGGCATTTTCGGTCATAATCA | |
| 2[47] | ACGGCTACAAAAGGAGCCTTTAATGTGAGAAT | |
| 22[239] | TTAACACCAGCACTAACAACTAATCGTTATTA | |
| 11[32] | AACAGTTTTGTACCAAAAACATTTTATTTC | |
| 1[160] | TTAGGATTGGCTGAGACTCCTCAATAACCGAT | |
| 23[64] | AAAGCACTAAATCGGAACCCTAATCCAGTT | |
| 19[56] | TACCGAGCTCGAATTCGGGAAACCTGTCGTGCAGCTGATT | |
| 19[160] | GCAATTCACATATTCCTGATTATCAAAGTGTA | |
| 15[32] | TAATCAGCGGATTGACCGTAATCGTAACCG | |
| 17[192] | CATTTGAAGGCGAATTATTCATTTTTGTTTGG | |
| 11[256] | GCCTTAAACCAATCAATAATCGGCACGCGCCT | |
| 23[224] | GCACAGACAATATTTTTGAATGGGGTCAGTA | |
| 0[239] | AGGAACCCATGTACCGTAACACTTGATATAA | |
| 9[64] | CGGATTGCAGAGCTTAATTGCTGAAACGAGTA | |
| 0[143] | TCTAAAGTTTTGTCGTCTTTCCAGCCGACAA | |
| 4[239] | GCCTCCCTCAGAATGGAAAGCGCAGTAACAGT | |
| 16[207] | ACCTTTTTATTTTAGTTAATTTCATAGGGCTT | |
| 20[111] | CACATTAAAATTGTTATCCGCTCATGCGGGCC | |
| 8[239] | AAGTAAGCAGACACCACGGAATAATATTGACG | |
| 7[192] | ATACATACCGAGGAAACGCAATAAGAAGCGCATTAGACGG | |
| 20[239] | ATTTTAAAATCAAAATTATTTGCACGGATTCG | |
| 1[32] | AGGCTCCAGAGGCTTTGAGGACACGGGTAA | |
| 23[96] | CCCGATTTAGAGCTTGACGGGGAAAAAGAATA | |
| 8[143] | CTTTTGCAGATAAAAACCAAAATAAAGACTCC | |
| 12[239] | CTTATCATTCCCGACTTGCGGGAGCCTAATTT | |
| 13[256] | GTTTATCAATATGCGTTATACAAACCGACCGTGTGATAAA | |
| 12[111] | TAAATCATATAACCTGTTTAGCTAACCTTTAA | |
| 6[79] | TTATACCACCAAATCAACGTAACGAACGAG | |
| 13[224] | ACAACATGCCAACGCTCAACAGTCTTCTGA | |
| 17[160] | AGAAAACAAAGAAGATGATGAAACAGGCTGCG | |
| 11[160] | CCAATAGCTCATCGTAGGAATCATGGCATCAA | |
| 1[256] | CAGGAGGTGGGGTCAGTGCCTTGAGTCTCTGAATTTACCG | |
| 23[128] | AACGTGGCGAGAAAGGAAGGGAAACCAGTAA | |
| 4[175] | CACCAGAAAGGTTGAGGCAGGTCATGAAAG | |
| 19[224] | CTACCATAGTTTGAGTAACATTTAAAATAT | |
| 10[143] | CCAACAGGAGCGAACCAGACCGGAGCCTTTAC | |
| 23[160] | TAAAAGGGACATTCTGGCCAACAAAGCATC | |
| 15[192] | TCAAATATAACCTCCGGCTTAGGTAACAATTT | |
| 22[207] | AGCCAGCAATTGAGGAAGGTTATCATCATTTT | |
| 5[160] | GCAAGGCCTCACCAGTAGCACCATGGGCTTGA | |
| 21[256] | GCCGTCAAAAAACAGAGGTGAGGCCTATTAGT | |
| 2[207] | TTTCGGAAGTGCCGTCGAGAGGGTGAGTTTCG | |
| 15[96] | ATATTTTGGCTTTCATCAACATTATCCAGCCA | |
| 3[160] | TTGACAGGCCACCACCAGAGCCGCGATTTGTA | |
| 11[64] | GATTTAGTCAATAAAGCCTCAGAGAACCCTCA | |
| 20[175] | ATTATCATTCAATATAATCCTGACAATTAC | |
| 8[79] | AATACTGCCCAAAAGGAATTACGTGGCTCA | |
| 4[143] | TCATCGCCAACAAAGTACAACGGACGCCAGCA | |
| 22[111] | GCCCGAGAGTCCACGCTGGTTTGCAGCTAACT | |
| 3[128] | AGCGCGATGATAAATTGTGTCGTGACGAGA | |
| 0[111] | TAAATGAATTTTCTGTATGGGATTAATTTCTT | |
| 7[96] | TAAGAGCAAATGTTTAGACTGGATAGGAAGCC | |
| 23[32] | CAAATCAAGTTTTTTGGGGTCGAAACGTGGA | |
| 8[111] | AATAGTAAACACTATCATAACCCTCATTGTGA | |
| 4[271] | AAATCACCTTCCAGTAAGCGTCAGTAATAA | |
| 16[143] | GCCATCAAGCTCATTTTTTAACCACAAATCCA | |
| 0[47] | AGAAAGGAACAACTAAAGGAATTCAAAAAAA | |
| 21[128] | GCGAAAAATCCCTTATAAATCAAGCCGGCG | |
| 18[175] | CTGAGCAAAAATTAATTACATTTTGGGTTA | |
| 0[207] | TCACCAGTACAAACTACAACGCCTAGTACCAG | |
| 3[32] | AATACGTTTGAAAGAGGACAGACTGACCTT | |
| 2[111] | AAGGCCGCTGATACCGATAGTTGCGACGTTAG | |
| 0[271] | CCACCCTCATTTTCAGGGATAGCAACCGTACT | |
| 9[32] | TTTACCCCAACATGTTTTAAATTTCCATAT | |
| 9[96] | CGAAAGACTTTGATAAGAGGTCATATTTCGCA | |
| 14[79] | GCTATCAGAAATGCAATGCCTGAATTAGCA | |
| 18[143] | CAACTGTTGCGCCATTCGCCATTCAAACATCA | |
| 18[207] | CGCGCAGATTACCTTTTTAATGGGAGAGACT | |
| 3[96] | ACACTCATCCATGTTACTTAGCCGAAAGCTGC | |
| 17[32] | TGCATCTTTCCCAGTCACGACGGCCTGCAG | |
| 7[128] | AGACGACAAAGAAGTTTTGCCATAATTCGAGCTTCAA | |
| 8[271] | AATAGCTATCAATAGAAAATTCAACATTCA | |
| 10[79] | GATGGCTTATCAAAAAGATTAAGAGCGTCC | |
| 19[32] | GTCGACTTCGGCCAACGCGCGGGGTTTTTC | |
| 19[128] | CACAACAGGTGCCTAATGAGTGCCCAGCAG | |
| 13[64] | TATATTTTGTCATTGCCTGAGAGTGGAAGATTGTATAAGC | |
| 5[32] | CATCAAGTAAAACGAACTAACGAGTTGAGA | |
| 1[192] | GCGGATAACCTATTATTCTGAAACAGACGATT | |
| 14[207] | AATTGAGAATTCTGTCCAGACGACTAAACCAA | |
| 9[224] | AAAGTCACAAAATAAACAGCCAGCGTTTTA | |
| 21[160] | TCAATATCGAACCTCAAATATCAATTCCGAAA | |
| 13[96] | TAGGTAAACTATTTTTGAGAGATCAAACGTTA | |
| 2[79] | CAGCGAAACTTGCTTTCGAGGTGTTGCTAA | |
| 22[271] | CAGAAGATTAGATAATACATTGTCGACAA | |
| 2[271] | GTTTTAACTTAGTACCGCCACCCAGAGCCA | |
| 20[143] | AAGCCTGGTACGAGCCGGAAGCATAGATGATG | |
| 16[79] | GCGAGTAAAAATATTTAAATTGTTACAAAG | |
| 10[175] | TTAACGTCTAACATAAAAACAGGTAACGGA | |
| 18[271] | CTTTTACAAAATCGTCGCTATTAGCGATAG | |
| 9[256] | GAGAGATAGAGCGTCTTTCCAGAGGTTTTGAA | |
| 20[79] | TTCCAGTCGTAATCATGGTCATAAAAGGGG | |
| 12[143] | TTCTACTACGCGAGCTGAAAAGGTTACCGCGC | |